\documentclass[preprint,12pt]{elsarticle}

\usepackage{amssymb}
\usepackage{amsmath}
\usepackage{comment}
\usepackage{siunitx}
\usepackage{subfig}
\usepackage{booktabs}
\usepackage{makecell}

\journal{NIMA}

\begin{document}

\begin{frontmatter}



\title{SiPM non-linearity studies in beam tests with scintillating crystals}


\author[a,b,c]{Zhiyu Zhao}
\author[d,e]{Dejing Du}
\author[a,b,c]{Shu Li}
\author[d,e]{Yong Liu\corref{cor1}}
\ead{liuyong@ihep.ac.cn}
\author[d,e]{Baohua Qi}
\author[d,e]{Jack Rolph}
\author[b,c,a] {Haijun Yang}

\cortext[cor1]{Corresponding authors.}

\address[a]{Tsung-Dao Lee Institute,
Shanghai Jiao Tong University,\\1 Lisuo Road, Shanghai 201210, China}
\address[b]{Institute of Nuclear and Particle Physics, School of Physics and Astronomy,\\800 Dongchuan Road, Shanghai 200240, China}
\address[c]{Key Laboratory for Particle Astrophysics and Cosmology (MOE), Shanghai Key Laboratory for Particle Physics and Cosmology (SKLPPC), Shanghai Jiao Tong University,\\800 Dongchuan Road, Shanghai 200240, China}
\address[d]{Institute of High Energy Physics, Chinese Academy of Sciences, 19B Yuquan Road, Beijing 100049, China}
\address[e]{University of Chinese Academy of Sciences, 19A Yuquan Road, Beijing 100049, China}

\begin{abstract}

High-granularity homogeneous electromagnetic calorimeters based on scintillating crystals and silicon photomultipliers (SiPMs) are a promising option for future $e^{+}e^{-}$ Higgs factories, where both excellent energy resolution and a very large dynamic range are required. In this work, the non-linear response of high-pixel-density SiPMs with pixel pitches of 6--10~$\mu$m coupled to BGO and BSO crystals is studied under realistic beam conditions. A dual-end readout scheme with an attenuated reference SiPM was employed to precisely calibrate the deposited energy and the corresponding number of photoelectrons over a wide dynamic range. Beam tests were carried out at the CERN SPS H2 beamline using high-energy electrons, with a tungsten pre-shower and variable incident angles to enhance energy deposition. The measurements directly quantify the non-linear response of SiPMs to scintillation light over an extended dynamic range. For BGO-coupled Hamamatsu SiPMs, deviations from linearity of about 20\% are observed at $5\times10^{5}$ photoelectrons, while larger deviations are measured for the tested NDL devices and for configurations with faster BSO scintillation.
\end{abstract}




\end{frontmatter}



\section{Introduction}
\label{intro}

A novel concept of high-granularity homogeneous calorimetry~\cite{Liu_2020,instruments6030040,qi2026} has been proposed for future Higgs factories, such as the Circular Electron–Positron Collider (CEPC)~\cite{CEPCTDR-Acc,CEPCTDR-Det}, which is designed to operate at centre-of-mass energies ranging from 91 to 360~GeV. Precision measurements of the Higgs boson properties, as well as those of the W and Z bosons, place stringent requirements on detector performance, in particular on energy resolution and spatial granularity, in order to fully exploit particle-flow algorithms. For electromagnetic calorimetry, dense and bright scintillating crystals, such as bismuth germanate (BGO)~\cite{JI2014143} and bismuth silicate (BSO)~\cite{ISHII2002201}, represent attractive candidates owing to their high density, moderate light yield, and mature mass-production technology. These properties enable the construction of compact calorimeter systems with fine segmentation.

The use of silicon photomultipliers (SiPMs)~\cite{PIEMONTE20192,KLANNER201936} as the photosensors for a homogeneous crystal calorimeter in a collider environment introduces several specific challenges. One of the most critical is the requirement to cover a very large dynamic range, spanning from the detection of signals at the level of minimum ionizing particles and hadronic $\tau$ decays, up to the precise measurement of electromagnetic showers induced by high-energy electrons and photons with energies approaching 180~GeV~\cite{CEPCTDR-Det}. Owing to the finite number of microcells in a SiPM, the device response becomes intrinsically non-linear when the incident scintillation light intensity approaches the saturation limit defined by the total pixel count.

Unlike SiPM applications in fields such as medical imaging, the requirements of a homogeneous crystal calorimeter at a Higgs factory are unprecedented, as they demand the simultaneous achievement of a very large dynamic range and an energy resolution at the percent level. The operation of SiPMs coupled to BGO or BSO crystal scintillators therefore defines a unique regime that has not yet been systematically explored. In particular, the relatively slow scintillation processes of these crystals give rise to pronounced pixel recovery effects, which can play a beneficial role in mitigating saturation at the shower maximum. As a consequence, existing non-linearity correction models developed for fast scintillators cannot be directly applied and must be carefully validated and adapted. Moreover, benchtop measurements alone are insufficient to fully characterize the detector response under realistic beam conditions and in the context of an integrated calorimeter system.

A key and advantageous feature of BGO and BSO crystals is their scintillation decay time. The pixel recovery effect, whereby a fired microcell regains sensitivity after its recharge time (typically 10--100~ns), allows a single pixel to be activated multiple times within a single scintillation pulse. For a slow scintillator such as BGO, with a decay time of approximately 300~ns, individual pixels can recover and be re-triggered several times over the duration of the light emission. This mechanism effectively increases the number of available pixels and significantly extends the dynamic range of the SiPM beyond its physical microcell count. Such a synergy is particularly advantageous for the electromagnetic showers expected at an $e^+e^-$ collider, as it implies that the effective saturation point of a BGO/BSO--SiPM system is substantially higher than what would be inferred from the nominal pixel number alone. Consequently, the slow scintillation decay of these crystals directly alleviates saturation effects and enhances the impact of pixel recovery, making BGO and BSO especially well suited for integration with SiPMs in the context of high-precision collider calorimetry.

This manuscript is organized as follows. Section~\ref{sec:design} introduces the experimental concept and the beam-test setup, including the simulation studies used to optimize the pre-shower configuration. Section~\ref{sec:calibration} describes the calibration chain, comprising the pre-amplifier charge-injection calibration, the SiPM gain calibration, the MIP-based energy calibration, and the relative light collection efficiency calibration. The beam-test results, including the reconstruction of deposited energy within crystals and the measured SiPM nonlinearity in terms of photoelectron counts for various crystal--SiPM configurations, are presented in Section~\ref{sec:beamtest}. The associated systematic uncertainties of the measurements are evaluated and discussed in Section~\ref{systematic_uncertainties}. Finally, the limitations of the present study and the outlook are summarized in Section~\ref{sec:discussions}.

\section{Experiment design and setup}
\label{sec:design}

The dynamic range of a SiPM is fundamentally limited by the number and density of its microcells. To study the non-linear response of ultra–high-dynamic-range SiPMs when detecting scintillation light from a crystal scintillator, it is necessary to induce sufficiently large energy depositions in the crystal so that the SiPM operates deep into its non-linear region. Equally important is the ability to calibrate, with high precision, the number of photons incident on the SiPM.

\begin{figure}[htbp]
    \centering
    \includegraphics[width=0.6\linewidth]{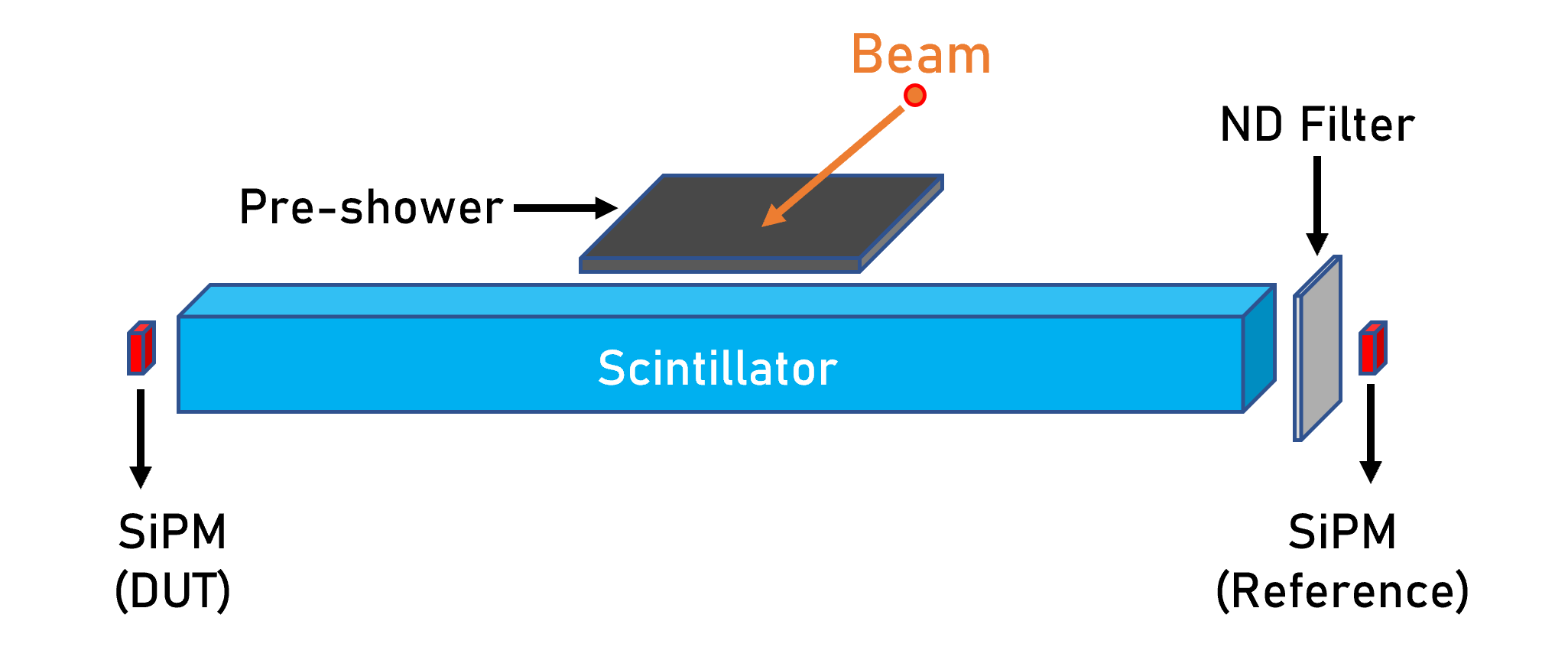}
    \caption{\label{fig:ini_design}~Schematic view of the experimental concept based on dual-end SiPM readout of a crystal bar.}
\end{figure}

To meet these requirements, an experimental concept based on dual-end SiPM readout of a crystal bar was developed, as illustrated in Figure~\ref{fig:ini_design}. A neutral-density (ND) filter was installed in front of the SiPM at one end of the crystal to attenuate the scintillation light, ensuring that this channel operates within its linear response region. This channel is referred to as $\mathrm{SiPM_{Ref}}$. The SiPM mounted at the opposite end, denoted as $\mathrm{SiPM_{DUT}}$ (device under test), received unattenuated scintillation light. In this configuration, the signal from $\mathrm{SiPM_{Ref}}$ was used to calibrate both the energy deposited in the crystal and the corresponding number of photons reaching $\mathrm{SiPM_{DUT}}$. Within the energy range of interest, the response of $\mathrm{SiPM_{Ref}}$ is therefore required to remain proportional to the deposited energy, which determines the choice of the ND filter transmittance.

To extend the measurement of SiPM non-linearity over a broad dynamic range, the energy deposition in the crystal was maximized by:
\begin{itemize}
  \item exploiting the high-energy electron beam (up to 300~GeV);
  \item installing a pre-shower absorber upstream of the crystal;
  \item steering the incident particles to traverse the crystal as parallel as possible to its longitudinal axis, while avoiding direct impacts on the SiPMs;
  \item optically coupling the crystal to $\mathrm{SiPM_{DUT}}$ using silicone oil to enhance the light collection efficiency.
\end{itemize}
The material and thickness of the pre-shower were optimized through dedicated simulation studies.

\subsection{Simulation studies for pre-shower}
\label{pre-shower_simulation}

The region of highest energy density in an electromagnetic shower is referred to as the shower maximum. In present study, the energy deposition in the crystal was maximized by placing an absorber layer upstream of the crystal to initiate shower development, while simultaneously rotating both the crystal and the pre-shower to increase the effective path length of the shower within the crystal. In addition to triggering shower development, the pre-shower is required to limit the transverse spread of the shower. Given the small transverse cross-section of the crystal bar, materials with short radiation lengths and small Molière radius are preferred, as they confine a larger fraction of the shower within the crystal volume and thereby enhance the absorbed energy.

\begin{table}[htbp]
\centering
\fontsize{9}{13}\selectfont
\caption{\label{tab:pre-shower}~Candidate materials considered for the pre-shower.~\cite{PDG}}
    \begin{tabular}{ccc}
        \toprule
        \makecell[c]{Material} & \makecell[c]{Radiation length (cm)} & \makecell[c]{Moli\`ere radius (cm)}\\ 
        \midrule
        W   & 0.35 & 0.93 \\
        Cu  & 1.44 & 1.57 \\
        BGO & 1.12 & 2.26 \\
        \bottomrule
    \end{tabular}
\end{table}

A Geant4\cite{Geant4}-based simulation was performed to compare the energy deposition in the crystal for three candidate pre-shower materials: tungsten (W), copper (Cu), and BGO. The relevant material properties are summarized in Table~\ref{tab:pre-shower}.

\begin{figure}[htbp]
    \centering
    \includegraphics[width=0.6\linewidth]{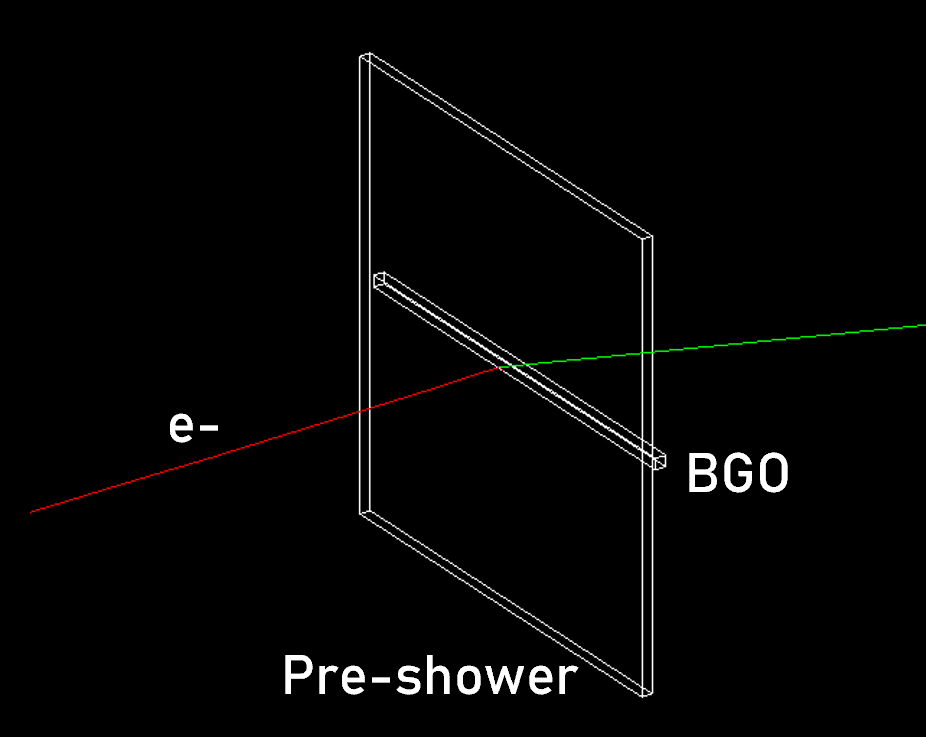}
    \caption{\label{fig:G4_pre-shower}~Geant4 simulation setup used for the pre-shower optimization studies.}
\end{figure}

\begin{figure}[htbp]
    \centering  
    \subfloat[]{
    \includegraphics[width=0.32\textwidth]{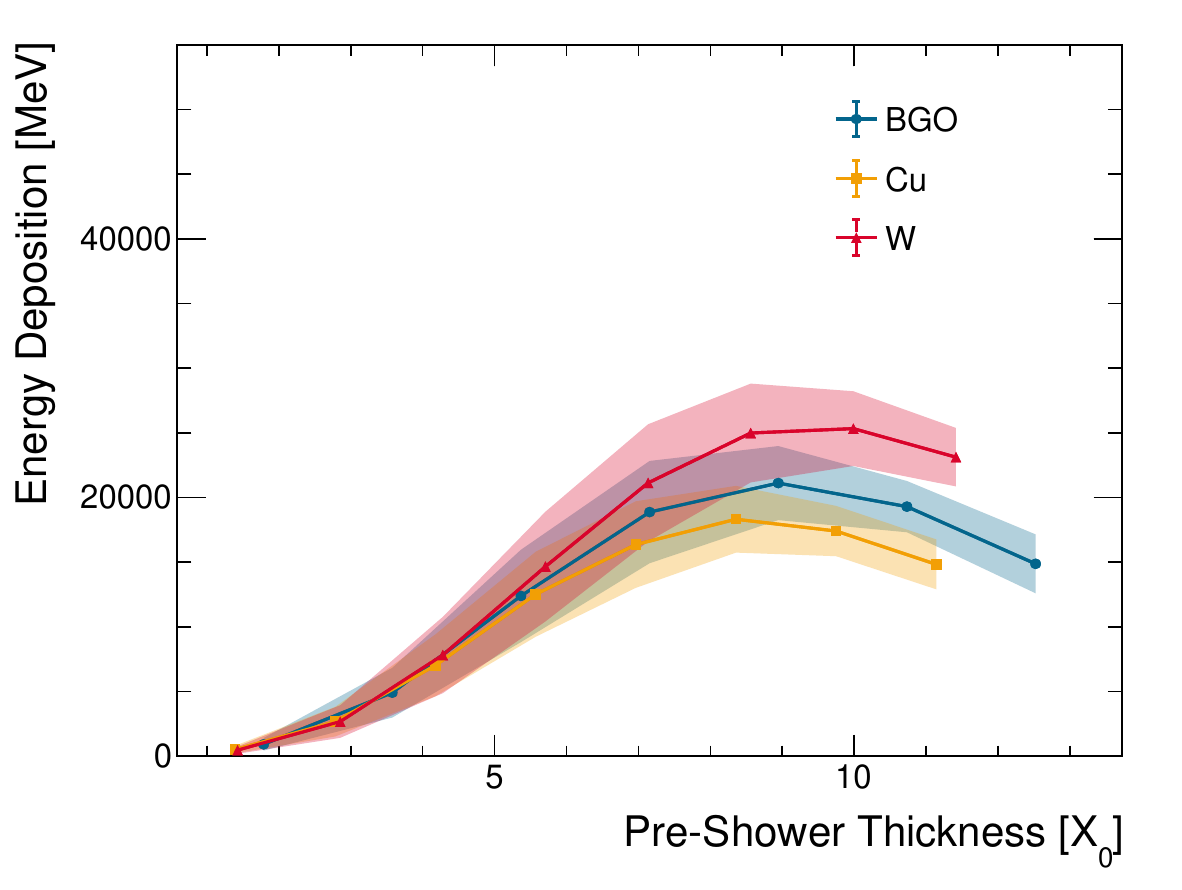}} 
    \subfloat[]{
    \includegraphics[width=0.32\textwidth]{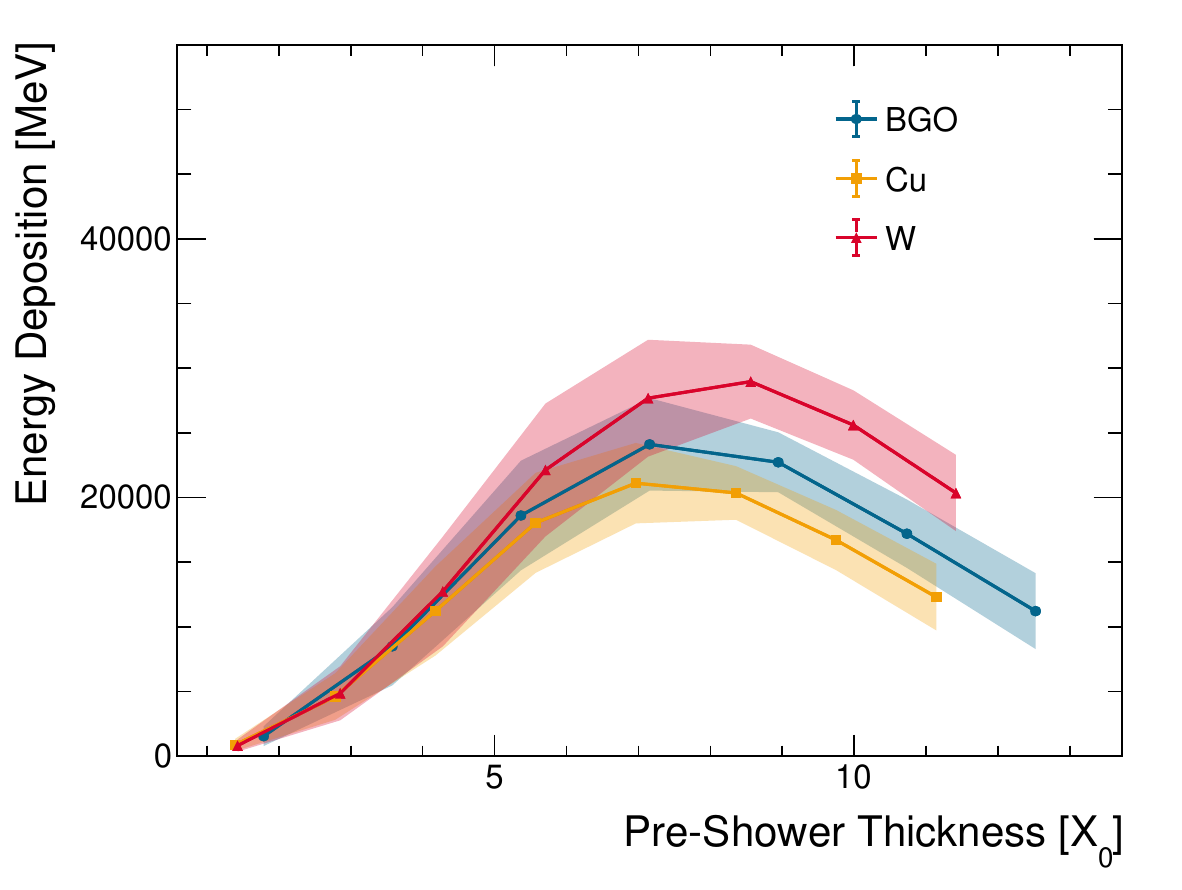}} 
    \subfloat[]{
    \includegraphics[width=0.32\textwidth]{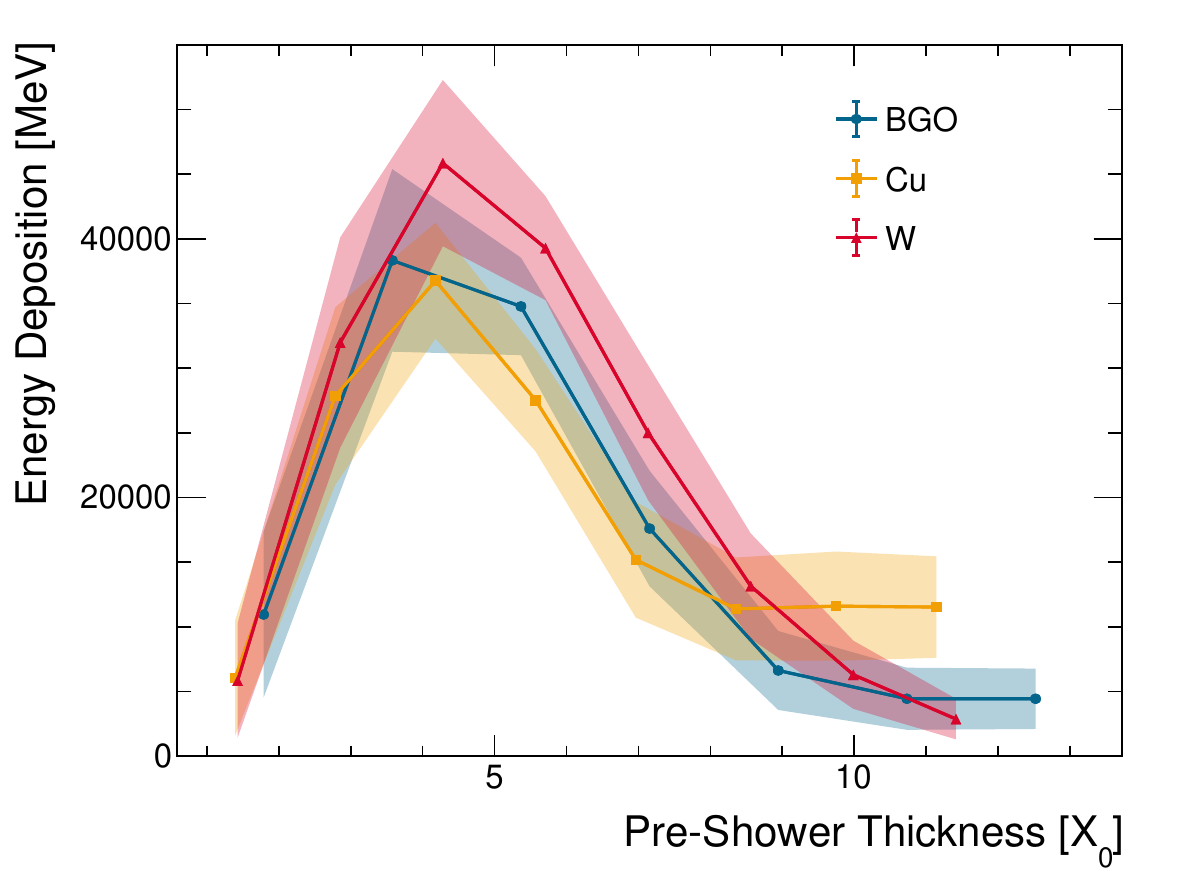}} 
    \caption{\label{fig:MC_BGO_Edep_angle}~Energy absorbed in the crystal as a function of pre-shower thickness for different incident angles: (a) $90^\circ$, (b) $60^\circ$, and (c) $30^\circ$ with respect to the crystal longitudinal axis.}
\end{figure}

In the simulation, a planar absorber was placed upstream of the crystal bar as the pre-shower (Figure~\ref{fig:G4_pre-shower}), and its thickness was varied while keeping the relative geometry between the absorber and the crystal fixed. Figures~\ref{fig:MC_BGO_Edep_angle} show the dependence of the energy absorbed by the crystal on the pre-shower thickness for particles incident at different angles with respect to the crystal longitudinal axis. As the absorber thickness increases, the absorbed energy initially rises, reaches a maximum, and then decreases. Among the three materials studied, tungsten yields the highest energy deposition in the crystal, which can be attributed to its small radiation length and Moli\`ere radius.

A comparison of Figures~\ref{fig:MC_BGO_Edep_angle} indicates that decreasing the angle between the particle trajectory and the crystal longitudinal axis significantly enhances the maximum absorbed energy, while simultaneously reducing the optimal pre-shower thickness. For instance, at an incident angle of $30^\circ$, the absorbed energy is approximately a factor of two larger than that at $90^\circ$, and the corresponding optimal absorber thickness is reduced by about a factor of two. Tungsten is therefore favored as the pre-shower material. 


\subsection{Experimental setup}
\label{Exp_setup}

\begin{figure}[htbp]
    \centering  
    \subfloat[]{
    \includegraphics[width=0.4\textwidth]{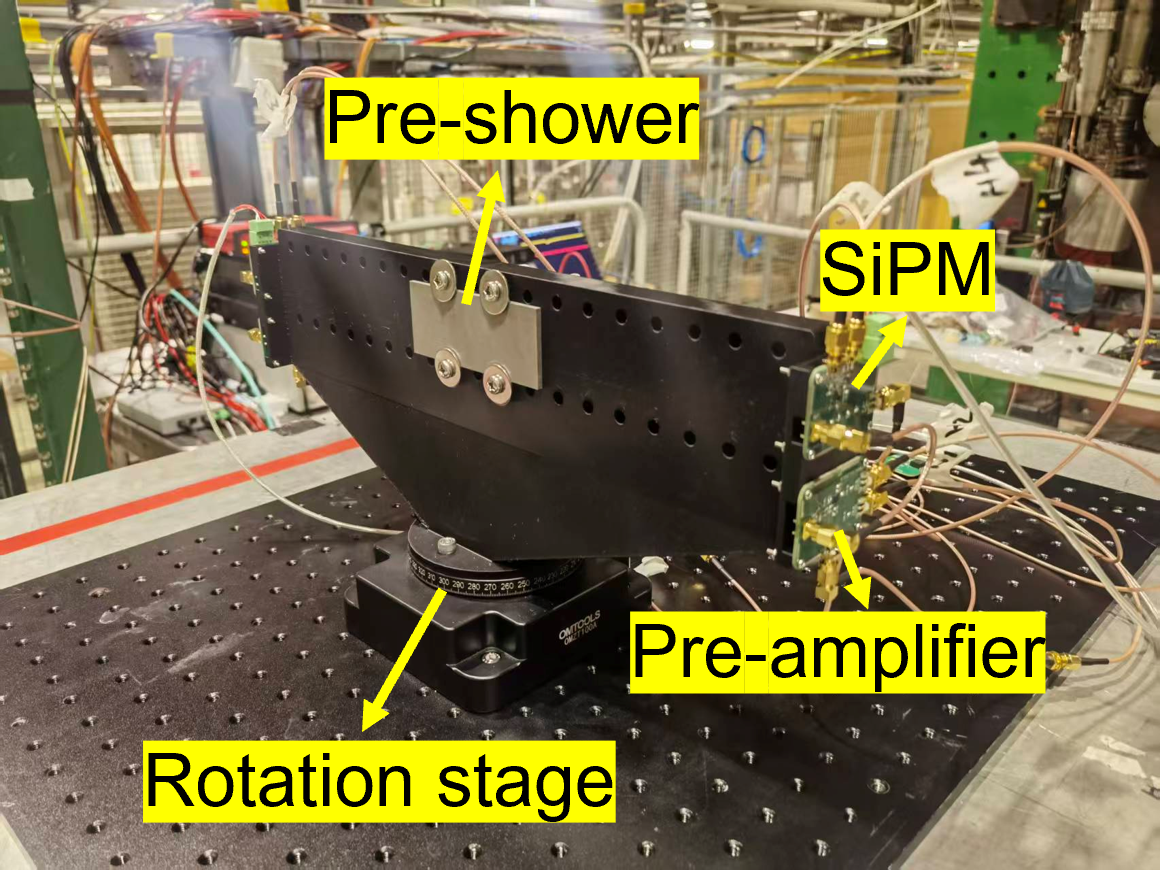}} 
    \subfloat[]{
    \includegraphics[width=0.5\textwidth]{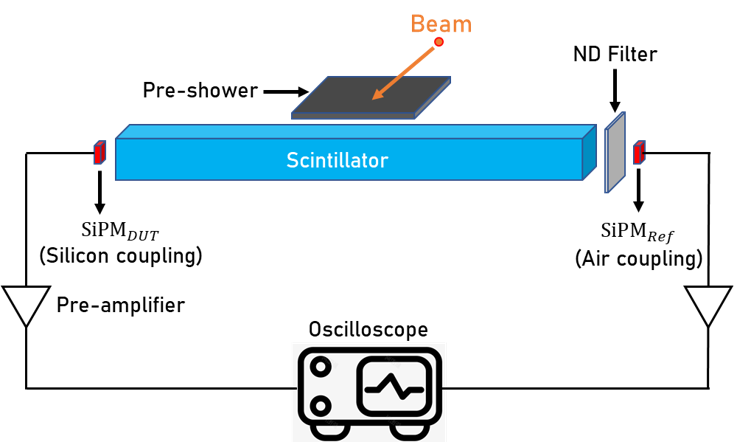}} 
    \caption{\label{fig:crystal_sipm}~Photograph and schematic view of the experimental setup.}
\end{figure}

Based on the simulation results, tungsten was selected as the pre-shower material. A motorized rotation stage was employed to control the angle between the crystal and the incident particle direction, enabling measurements over a wide dynamic range of deposited energies. The experimental setup is shown in Figure~\ref{fig:crystal_sipm}. It consisted of a crystal bar, a tungsten pre-shower plate, a neutral-density (ND) filter, an automated rotation platform, front-end electronics boards (including the SiPM readout and pre-amplifier boards), a high-speed oscilloscope, and a custom 3D-printed PLA mechanical support. The measurements were performed at the CERN SPS H2 beamline using 300~GeV electrons for electromagnetic shower studies and 160~GeV pions for minimum ionizing particle (MIP) energy calibration. 

\subsubsection{Crystal scintillators and SiPMs}
\label{crystal_SiPM}

Three crystal bars and four different SiPM models with pixel sizes as small as $6$--$10~\mu\mathrm{m}$ were tested, as summarized in Table~\ref{tab:crystals} and \ref{tab:SiPMs}. For the $40\times1.5\times1.5~\mathrm{cm}^3$ BGO crystal bar, all four SiPM types were evaluated. The other two, shorter BGO and BSO crystal bars were tested only with the HPK S14160-6010PS SiPM.

Three crystal bars and four different SiPM models with pixel pitches ranging from $6$ to $10~\mu\mathrm{m}$ were investigated, as summarized in Tables~\ref{tab:crystals} and~\ref{tab:SiPMs}. For the $40\times1.5\times1.5~\mathrm{cm}^3$ BGO crystal bar, all four SiPM types were tested. The two shorter crystal bars, one BGO and one BSO, both with dimensions of $12\times2\times2~\mathrm{cm}^3$, were evaluated using the Hamamatsu S14160-3010PS SiPM.

\begin{table}[h]
\centering
\fontsize{8}{12}\selectfont
\caption{\label{tab:crystals}~Properties of the crystal scintillators used in this study.}
    \begin{tabular}{cccc}
        \toprule
        \makecell[c]{Crystal material} & \makecell[c]{Dimensions (cm$^3$)} & \makecell[c]{Light yield (ph/MeV)} & \makecell[c]{Decay time (ns)}\\ 
        \midrule
        BGO & $40\times1.5\times1.5$ & 8200 & 300 / 60 \\
        BGO & $12\times2\times2$     & 8200 & 300 / 60 \\
        BSO & $12\times2\times2$     & 2000 & 100 / 25 \\
        \bottomrule
    \end{tabular}
\end{table}

\begin{table}[htbp]
\centering
\fontsize{8}{12}\selectfont
\caption{\label{tab:SiPMs}~Key parameters of the tested SiPMs.}
    \begin{tabular}{ccccc}
        \toprule
        \makecell[c]{SiPM} &\makecell[c]{Pixel pitch ($\mu$m)}  &\makecell[c]{Active area (mm$^2$)} &\makecell[c]{Number of pixels} &\makecell[c]{PDE (\%) \\ $\lambda=\lambda_p$}\\ 
        \midrule
        HPK S14160-3010PS~\cite{Hamamatsu} & 10 & 3.0$\times$3.0 & 89984 & 18\% \\
        HPK S14160-6010PS~\cite{Hamamatsu} & 10 & 6.0$\times$6.0 & 359011 & 18\% \\
        NDL EQR06 11-3030D-S~\cite{NDL} & 6 & 3.0$\times$3.0 & 90000 & 36\% \\
        NDL EQR10 11-3030D-S~\cite{NDL} & 10 & 3.0$\times$3.0 & 244719 & 30\% \\
        \bottomrule
    \end{tabular}
\end{table}


\subsubsection{Readout electronics}
\label{electronics}

\begin{figure}[htbp]
    \centering  
    \subfloat[]{
    \includegraphics[width=0.42\textwidth]{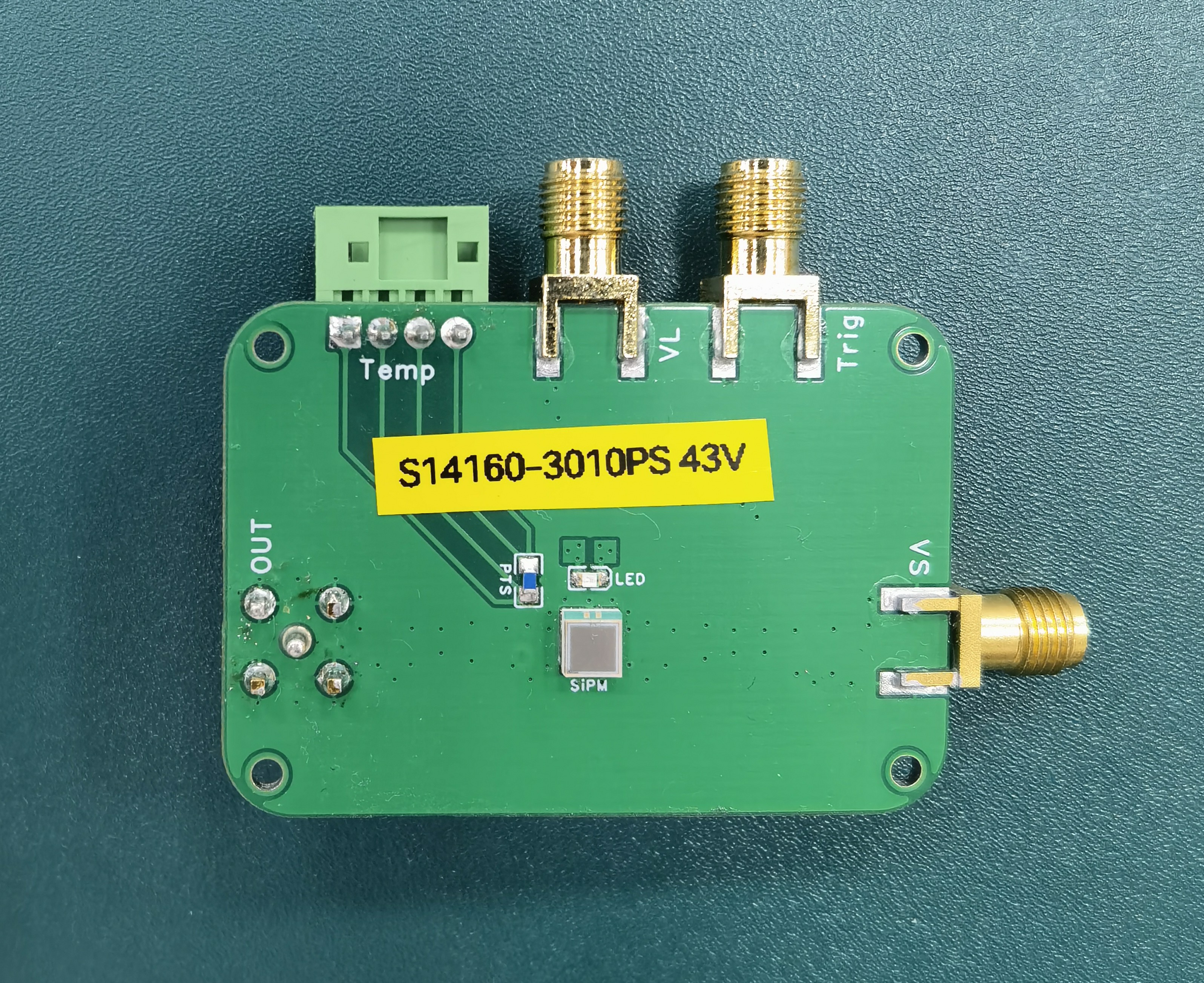}} 
    \subfloat[]{
    \includegraphics[width=0.446\textwidth]{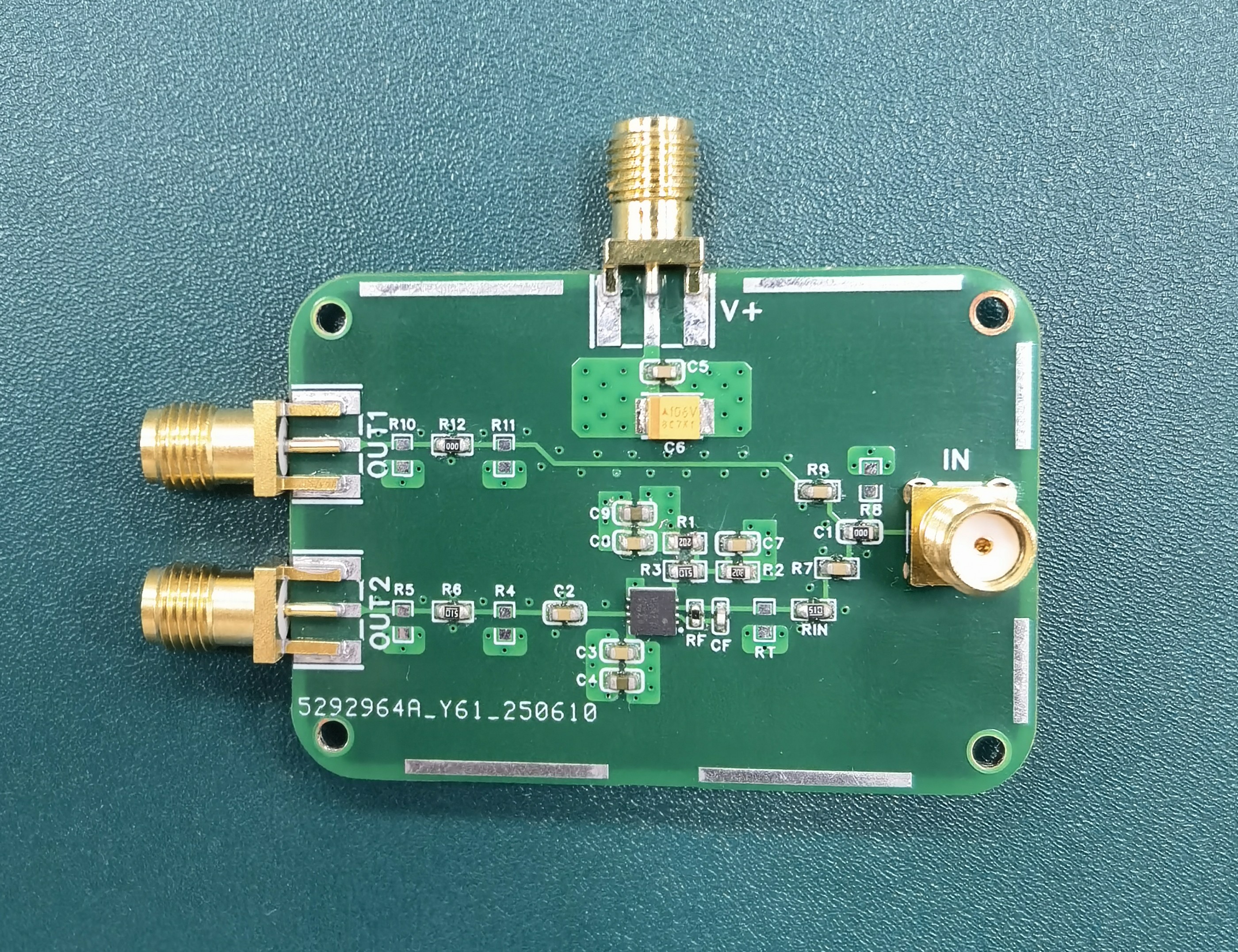}} 
    \caption{\label{fig:electronics}~Photographs of the SiPM board and the pre-amplifier board.}
\end{figure}


The front-end electronics consisted of a dedicated SiPM readout board and a pre-amplifier board (Figure~\ref{fig:electronics}). The SiPMs were operated with cathode biasing, and the signals were read out from the anode. A platinum resistance thermometer was mounted in close proximity to each SiPM to monitor its operating temperature, and an LED with a dedicated driver circuit~\cite{Tang_2023} was integrated for gain calibration. The LED pulse width was adjustable and could be set to values below 10~ns. 

The pre-amplifier board was designed to provide dual-gain readout, enabling simultaneous high- and low-gain signal acquisition to cover a wide dynamic range. The SiPM signal was split into two paths: one path was read out directly as the low-gain channel, while the other was amplified by a transimpedance amplifier with a gain of ten and recorded as the high-gain channel. Both channels were digitized simultaneously using a PicoScope~6426E~\cite{Pico} oscilloscope with a bandwidth of 1~GHz and a sampling rate of 1.25~GS/s.

\subsubsection{Neutral-density filter}
\label{ND_filter}

A reflective neutral-density (ND) filter with a nominal transmittance of 1\% was inserted between one end of the crystal and the $\mathrm{SiPM_{Ref}}$. The opposite end was optically coupled to the SiPM using silicone oil in order to maximize the photon collection efficiency for $\mathrm{SiPM_{DUT}}$ and to enable the investigation of SiPM non-linearity at high photon intensities.

\subsubsection{Mechanics}
\label{mechanics}

\begin{figure}[htbp]
    \centering  
    \subfloat[]{
    \includegraphics[width=0.32\textwidth]{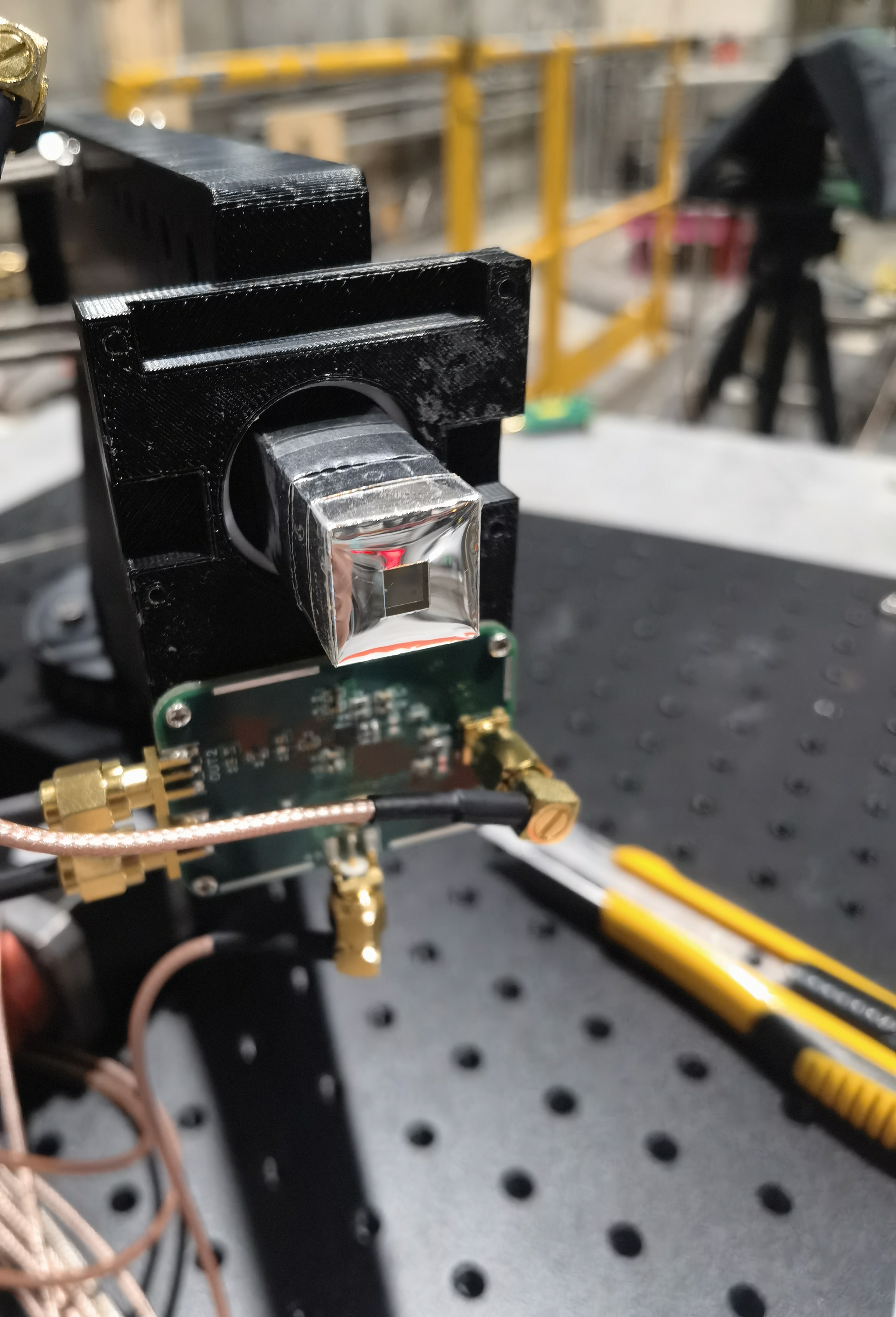}} 
    \subfloat[]{
    \includegraphics[width=0.32\textwidth]{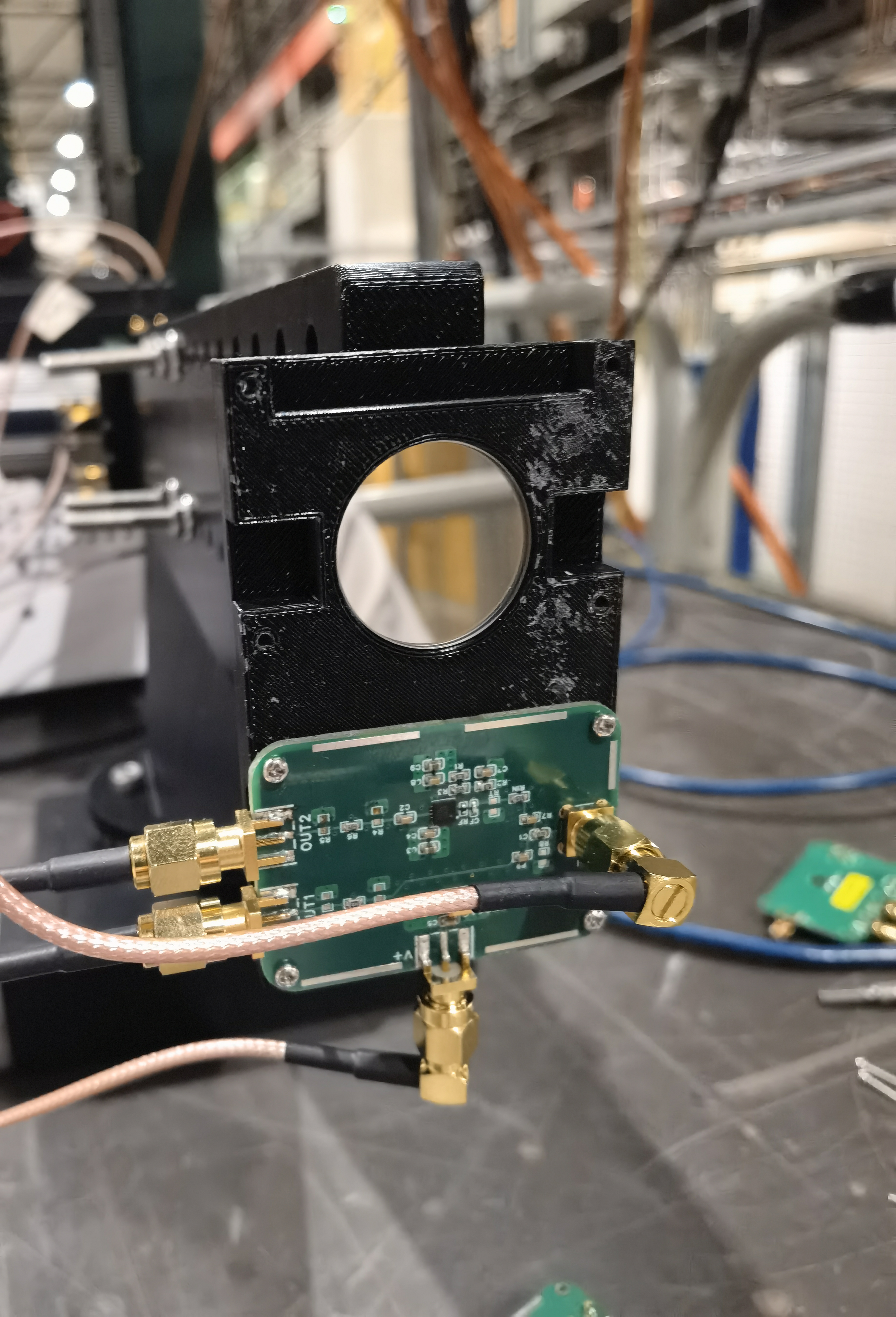}} 
    \subfloat[]{
    \includegraphics[width=0.32\textwidth]{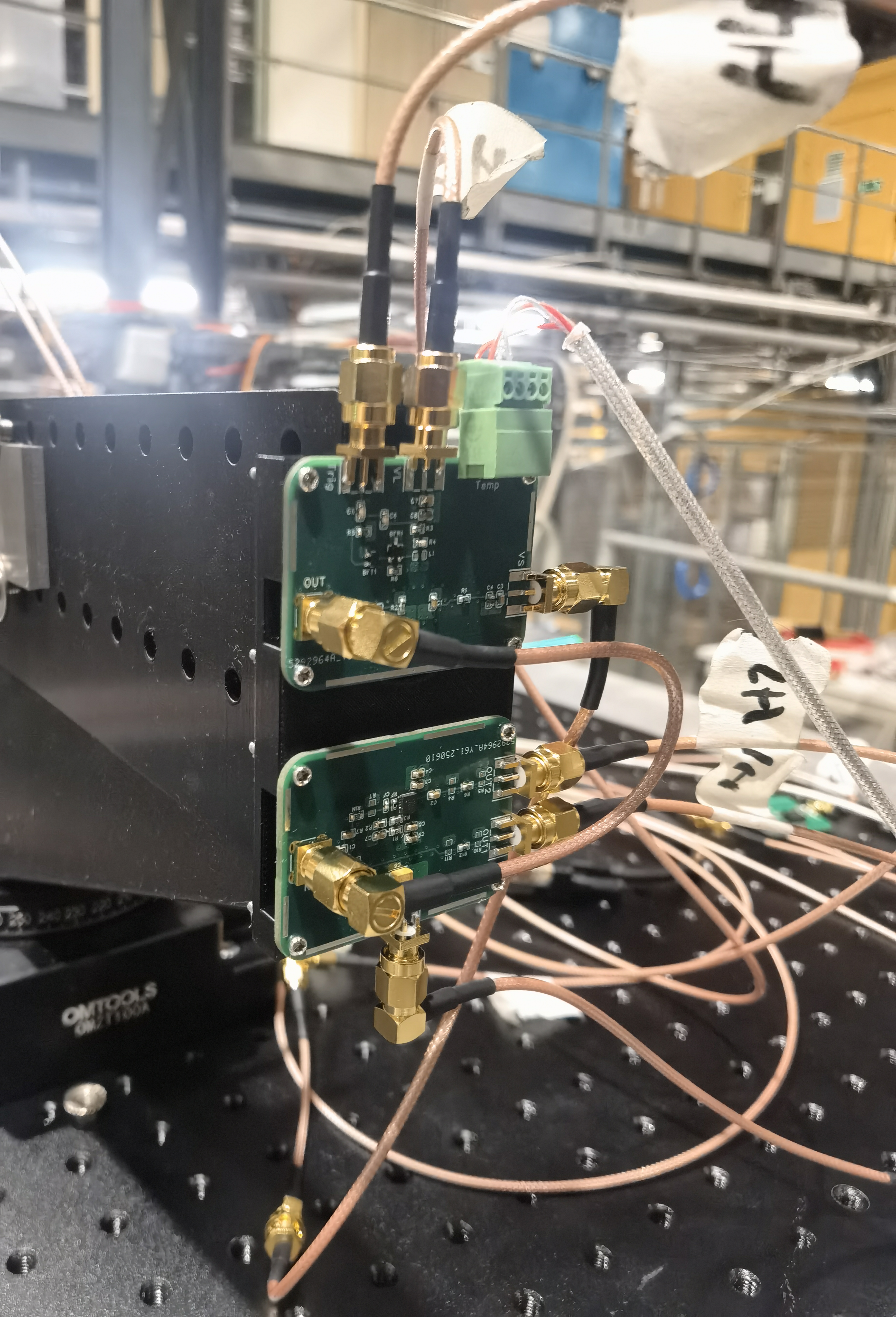}} 
    \caption{\label{fig:mechanics}~Mechanical details of the experimental setup.}
\end{figure}

As shown in Figure~\ref{fig:mechanics}, the crystal bar was housed in a PLA support structure. The tungsten pre-shower plate was mounted at the geometric center of the crystal front face using screws, with a separation of approximately 2~mm from the crystal surface. The front-end electronics boards (lower board) were attached to both ends of the crystal, providing a total of four readout channels. The complete assembly was fixed to a motorized rotation platform with full $360^\circ$ rotation capability, allowing remote adjustment of the incident angle. Prior to data taking, careful alignment of the beam and the experimental setup was performed to ensure that the particle beam was centered on the crystal and the pre-shower.







\section{Calibration}
\label{sec:calibration}

\begin{figure}[htbp]
    \centering  
    \subfloat[]{
    \includegraphics[width=0.45\textwidth]{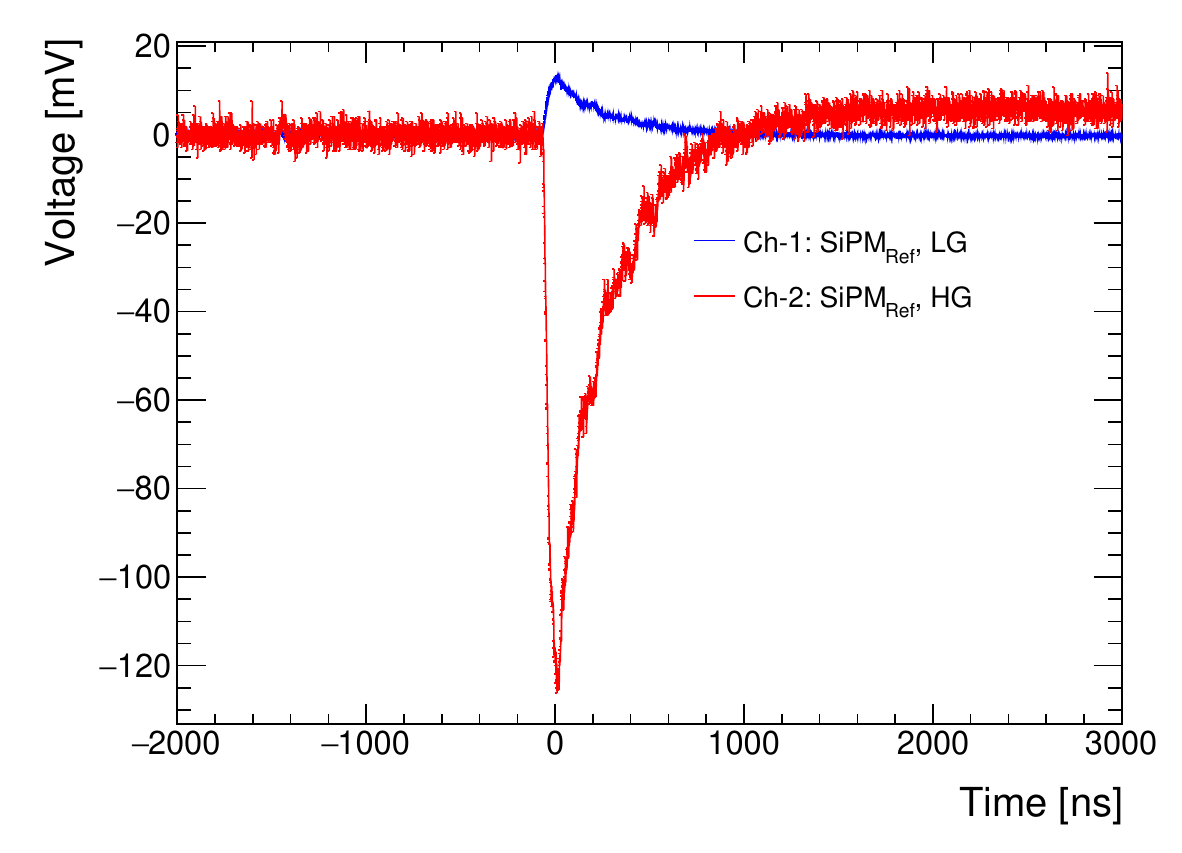}} 
    \subfloat[]{
    \includegraphics[width=0.45\textwidth]{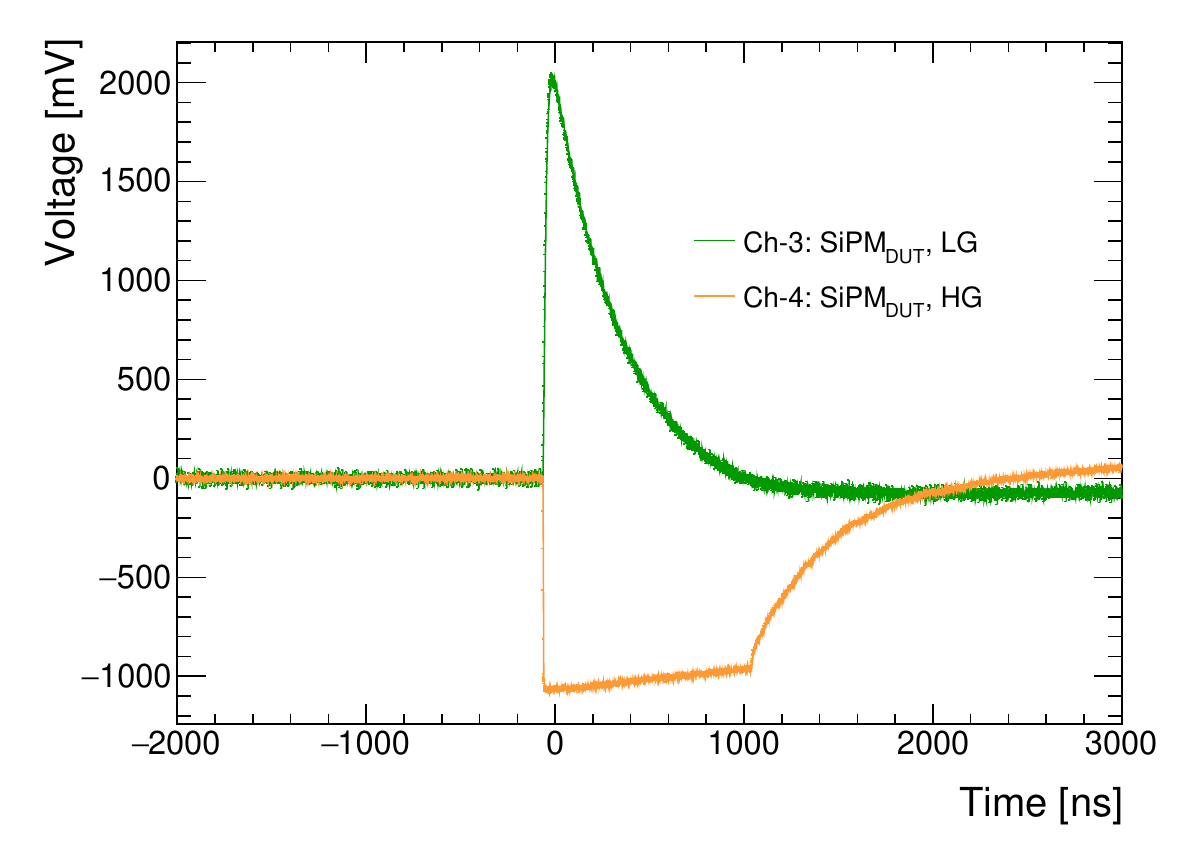}} 
    \caption{\label{fig:waveform}~Example waveforms measured in the beam test: (a) $\mathrm{SiPM_{Ref}}$ channels; (b) $\mathrm{SiPM_{DUT}}$ channels.}
\end{figure}

Typical waveforms recorded from the detector are shown in Figure~\ref{fig:waveform}. The readout provided four channels in total: both $\mathrm{SiPM_{DUT}}$ and $\mathrm{SiPM_{Ref}}$ were instrumented with a high-gain and a low-gain output. For each SiPM, a offline switching threshold was defined to select the appropriate gain channel on an event-by-event basis: the high-gain channel was used when its signal amplitude remained below the threshold, while the low-gain channel was used otherwise. Prior to applying the gain selection, a sequence of calibrations was performed to convert the digitized waveforms into deposited energy and photoelectron (p.e.) counts, and to enable a direct and quantitative evaluation of the SiPM response nonlinearity. The calibration procedure comprises four main steps: pre-amplifier calibration, SiPM gain calibration, relative light collection efficiency calibration, and MIP-based energy calibration.

\subsection{Pre-amplifier calibration via charge injection}
\label{Preamp_calib}

\begin{figure}[htbp]
    \centering
    \includegraphics[width=0.6\linewidth]{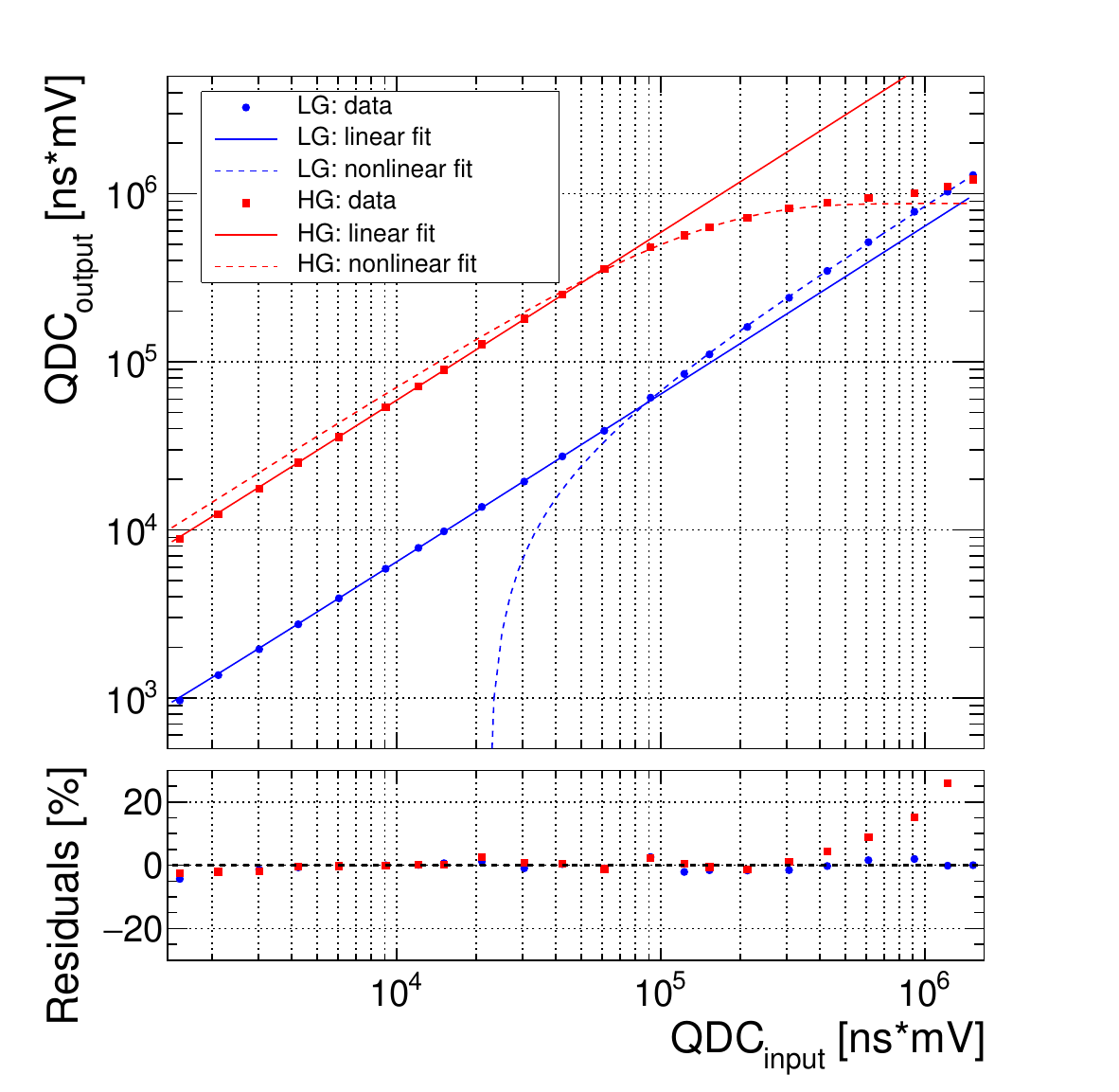}
    \caption{\label{fig:preamp_calib}~Charge-injection calibration of the pre-amplifier: output QDC versus input QDC for (a) the left-side electronics chain and (b) the right-side electronics chain.}
\end{figure}

The pre-amplifier transfer function is not strictly linear over the full dynamic range. The relationship between the injected input charge and the measured output charge for the dual-gain channels was therefore characterized using a charge-injection method. A waveform generator was used to emulate the typical crystal--SiPM pulse shape, and the generated pulses were injected into the pre-amplifier input. The input and output charges were quantified by the integral of the waveform (QDC). Figure~\ref{fig:preamp_calib} shows the measured output QDC as a function of the injected input QDC for pre-amplifier connected to $\mathrm{SiPM_{Ref}}$. Both the high- and low-gain channels exhibit an approximately linear response for input QDC values below $\sim6\times10^{4}\ \mathrm{ns\cdot mV}$, while deviations from linearity become visible at larger input signals.


A piecewise parameterization was used to describe the input--output relation: a linear function was applied in the small-signal region, while an exponential function was used for the large-signal region. The relative residuals of the fit are shown in the lower panel of Figure~\ref{fig:preamp_calib}. For subsequent analysis, only the small-signal region of the high-gain channel and the large-signal region of the low-gain channel are relevant for the gain-switching scheme; both regions are well described by the chosen parameterization. The fitted functions were inverted and applied to the beam-test data to correct the measured QDC values. 

\subsection{SiPM gain calibration}
\label{SiPM_gain_calib}

The SiPM gain was calibrated using short light pulses from an LED. The LED was positioned adjacent to the SiPM such that a fraction of the emitted light, after reflections, was detected by the SiPM. By adjusting the LED intensity, charge spectra with resolved photoelectron peaks were recorded and used to extract the single-photoelectron gain.


\begin{figure}[htbp]
    \centering
    \includegraphics[width=0.9\linewidth]{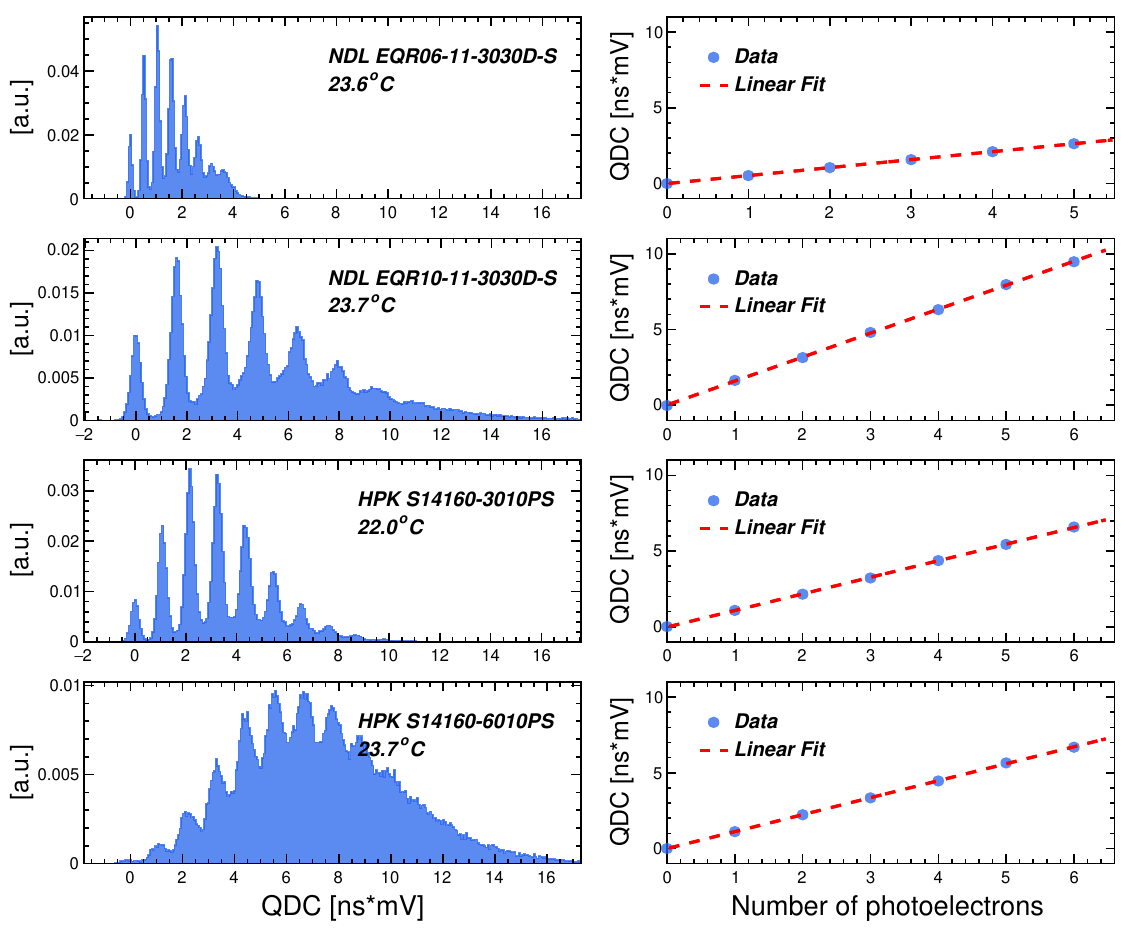}
    \caption{\label{fig:sipm_spectra}~Single-photoelectron charge spectra (left) and the corresponding QDC as a function of the number of photoelectrons (right) used for the SiPM gain calibration of the tested SiPM models at their respective beam test temperatures.
}
\end{figure}

The temperature dependence of the SiPM gain was taken into account by monitoring the local temperature with a PT100 platinum resistor mounted near each SiPM during the beam test. Dedicated offline measurements were subsequently carried out in a temperature-controlled chamber: the SiPMs were stabilized at temperatures matching those observed during the beam period, and the corresponding gains were measured. For some devices, like Hamamatsu S14160-6010PS, the gain at the beam-test temperature could not be determined directly because the combination of low gain and high dark-count rate prevented a clear single-photoelectron spectrum. Measurements were therefore performed at lower temperatures, where well-resolved spectra were obtained, and the gains were then extrapolated to the recorded temperatures. Representative charge spectra for the four SiPM models at their corresponding measurement temperatures are shown in Figure~\ref{fig:sipm_spectra}. Peak positions are extracted from the spectra on the left of Figure~\ref{fig:sipm_spectra} by fitting with multiple-Gaussian functions, which yields the QDC values corresponding to different photoelectron counts. A linear fit of the QDC versus number-of-photoelectrons distribution then gives the QDC per photoelectron - i.e., the SiPM gain.

\subsection{MIP energy calibration}
\label{MIP_calib}

A MIP energy calibration was performed for each detector configuration using $160~\mathrm{GeV}$ pion data. Beam particles were incident perpendicular to the geometric center of each crystal bar. The expected MIP energy deposition per unit path length was obtained from simulation and is summarized in Table~\ref{tab:MIP}. Here, the MIP energy is defined as the energy deposited by a particle traversing a reference length of the crystal, corresponding to 1.5~cm for the 40~cm-long BGO crystal and to 2.0~cm for the 12~cm-long BGO and BSO crystals.

\begin{table}[htbp]
\centering
\fontsize{9}{13}\selectfont
\caption{\label{tab:MIP}~Simulated MIP energy deposition for the crystals listed in Table~\ref{tab:crystals}.}
    \begin{tabular}{lc}
        \toprule
        \makecell[c]{Crystal} & \makecell[c]{MIP energy}\\ 
        \midrule
        $40\times1.5\times1.5~\mathrm{cm}^3$ BGO                           & 13.3~MeV/MIP \\
        $12\times2\times2~\mathrm{cm}^3$ BGO                               & 17.8~MeV/MIP \\
        $12\times2\times2~\mathrm{cm}^3$ BSO                               & 17.3~MeV/MIP \\
        \bottomrule
    \end{tabular}
\end{table}

The measured deposited-energy spectra are well described by a Landau distribution convolved with a Gaussian resolution function. The most probable value (MPV) of the convolution is taken as the reference MIP energy loss for calibration purposes.

\begin{figure}[htbp]
    \centering
    \includegraphics[width=0.6\linewidth]{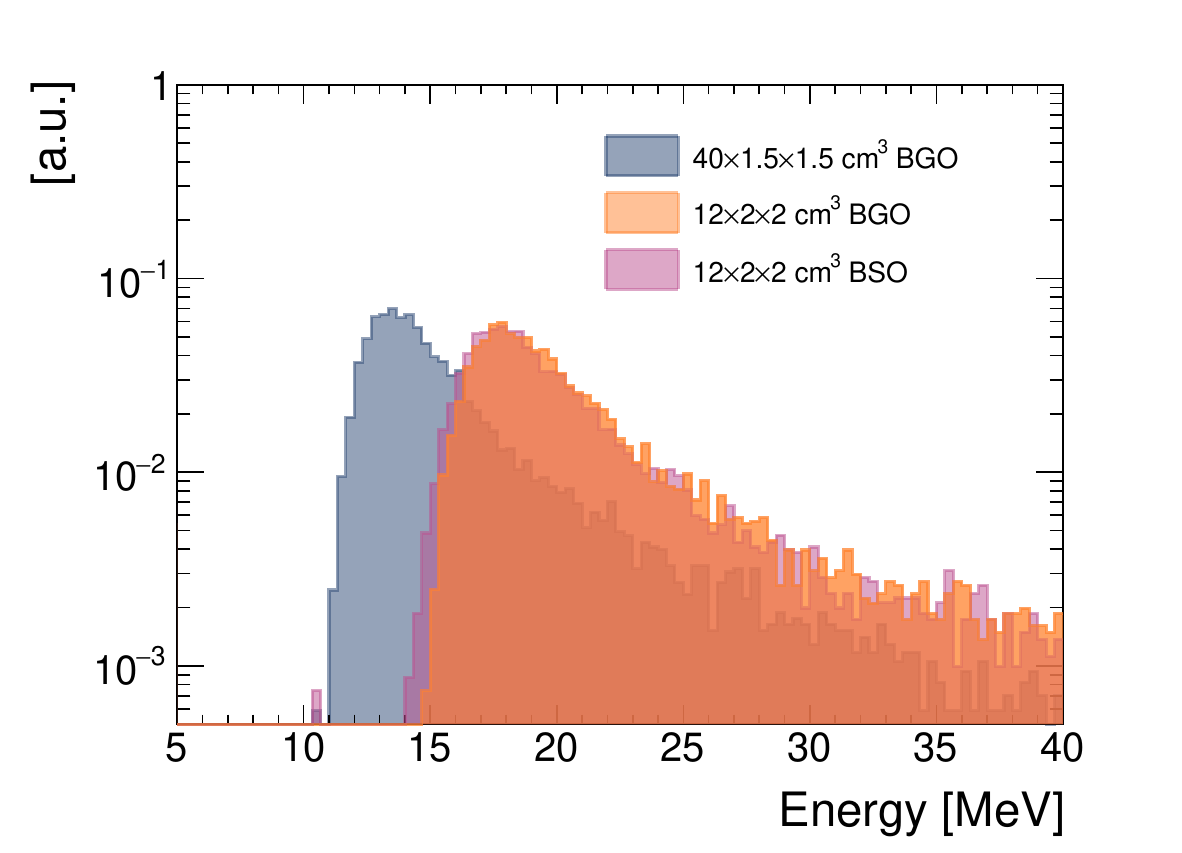}
    \caption{\label{fig:mip_spectrum}~Deposited energy spectrum after MIP calibration with $160~\mathrm{GeV}$ pions.}
\end{figure}

The experimental spectra were fitted with a Landau--Gaussian function to extract the MPV, which was subsequently used to determine the QDC-to-energy conversion factor (QDC per unit deposited energy) for each crystal--SiPM detection unit. Figure~\ref{fig:mip_spectrum} shows the calibrated deposited-energy spectra for crystal--SiPM units employing different crystal bars but the same SiPM model (Hamamatsu S14160-6010PS).

\subsection{Relative light collection efficiency calibration}
\label{sec:rel_LCE}

\begin{figure}[htbp]
    \centering  
    \subfloat[]{
    \includegraphics[width=0.32\textwidth]{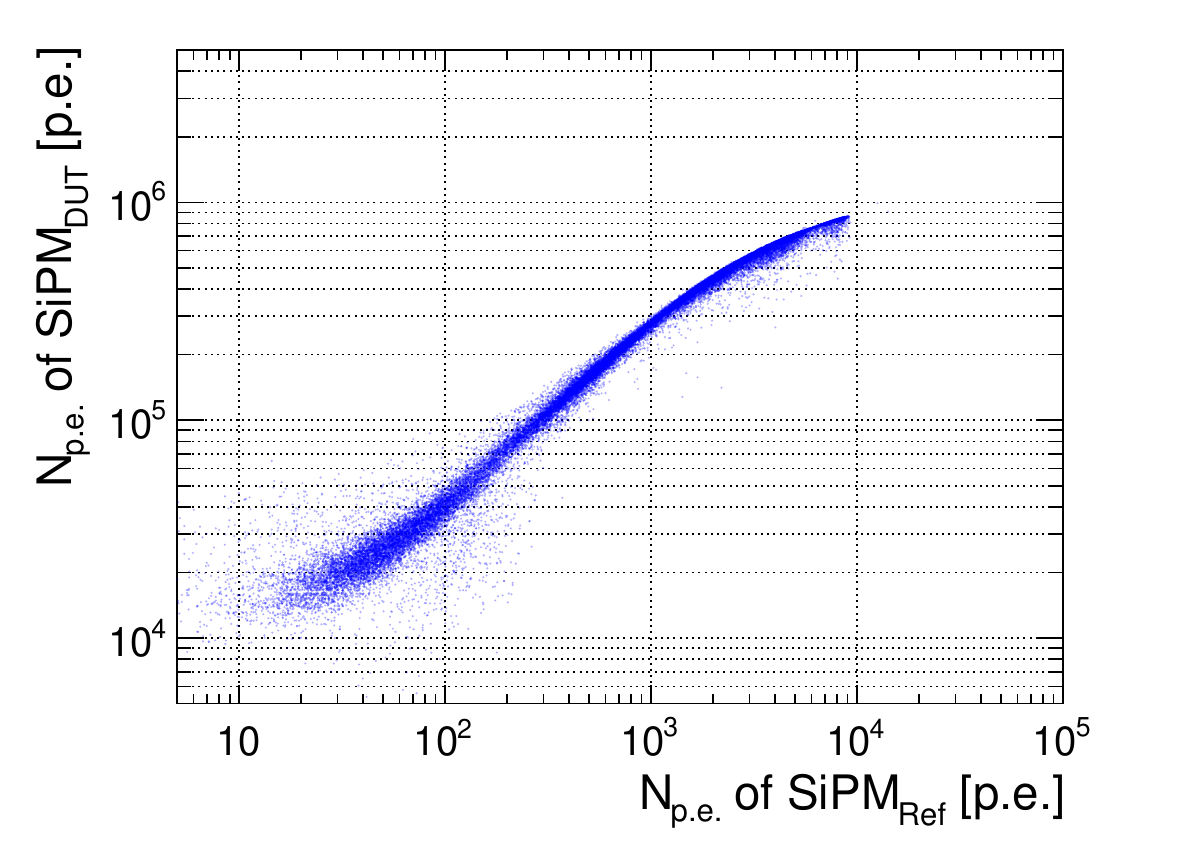}} 
    \subfloat[]{
    \includegraphics[width=0.32\textwidth]{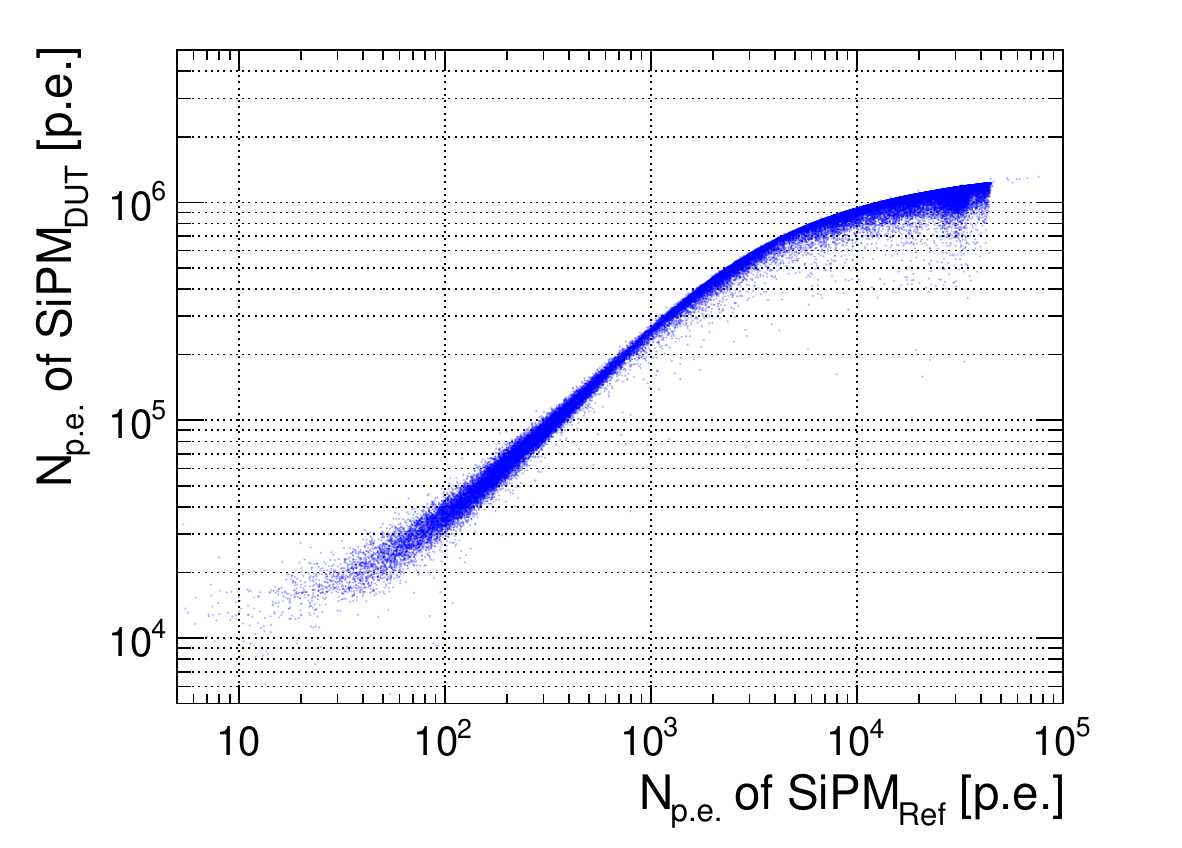}} 
    \subfloat[]{
    \includegraphics[width=0.32\textwidth]{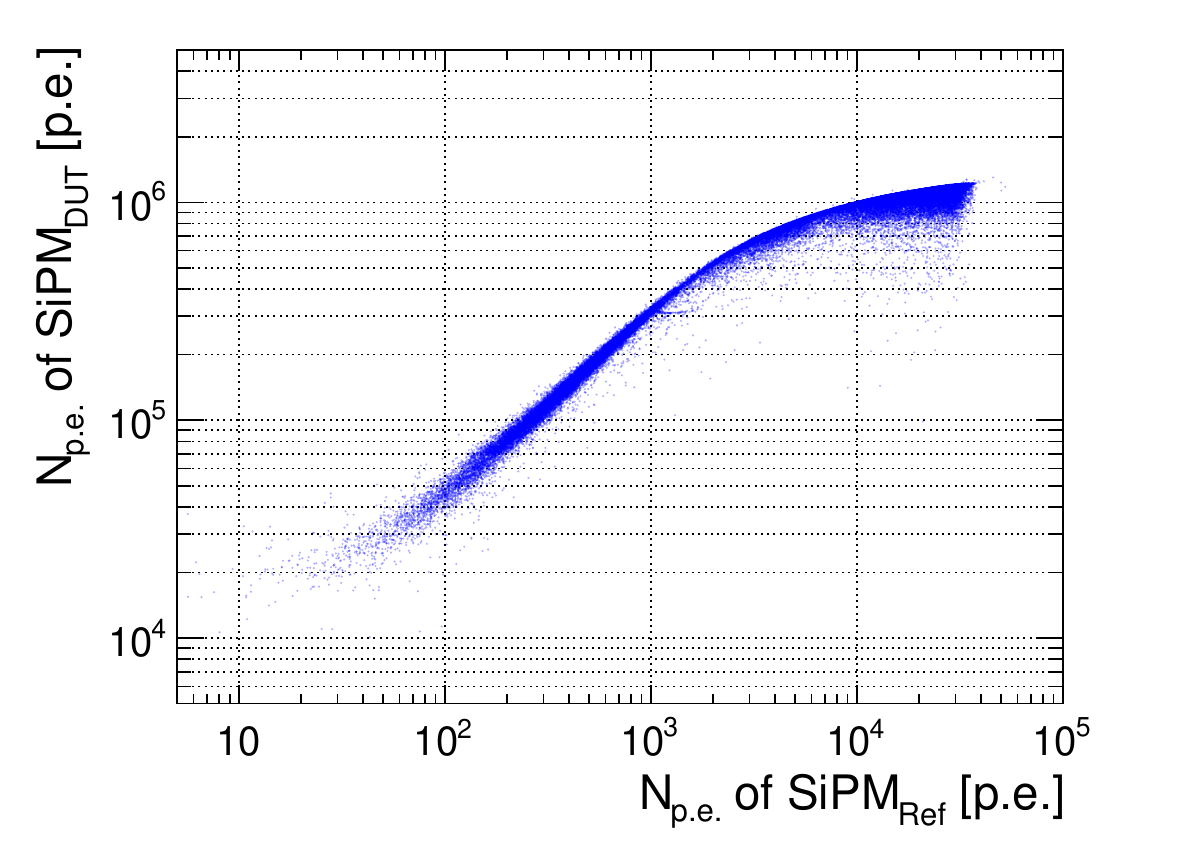}} \\
    \subfloat[]{
    \includegraphics[width=0.32\textwidth]{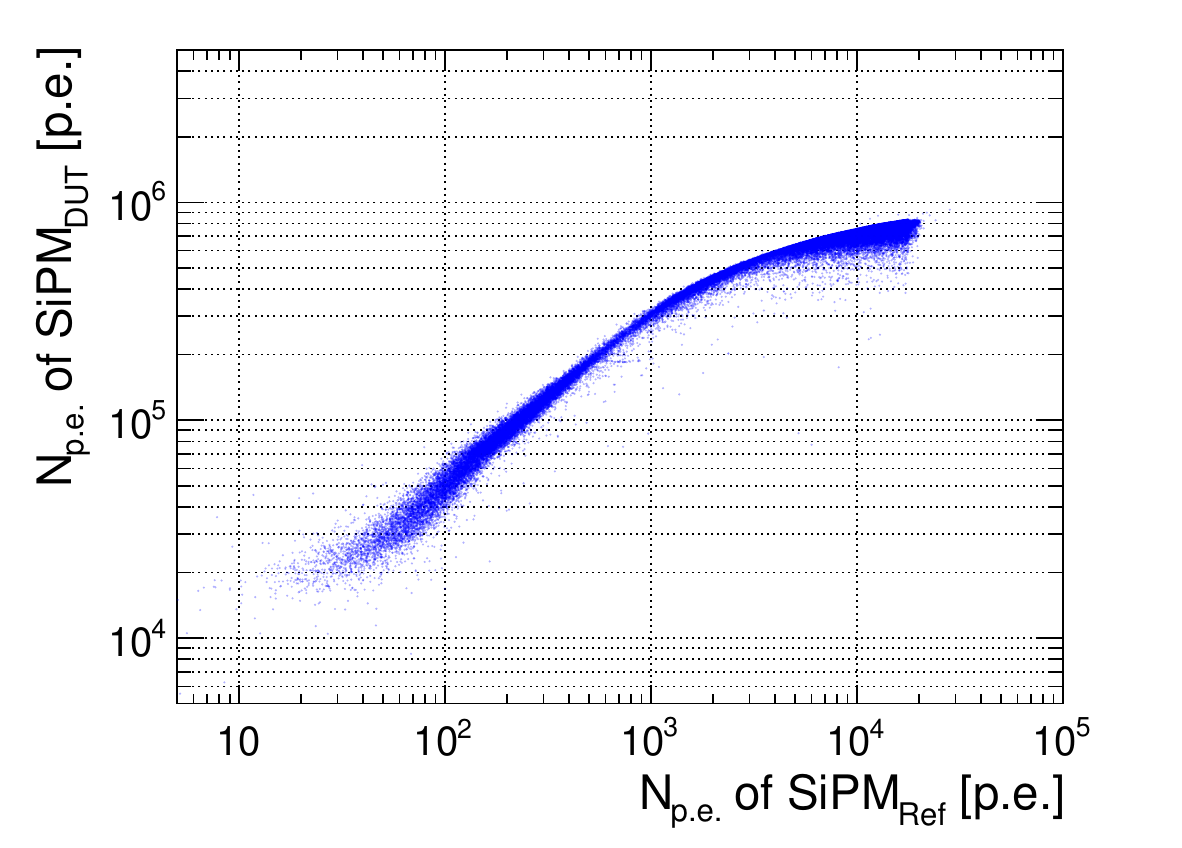}} 
    \subfloat[]{
    \includegraphics[width=0.32\textwidth]{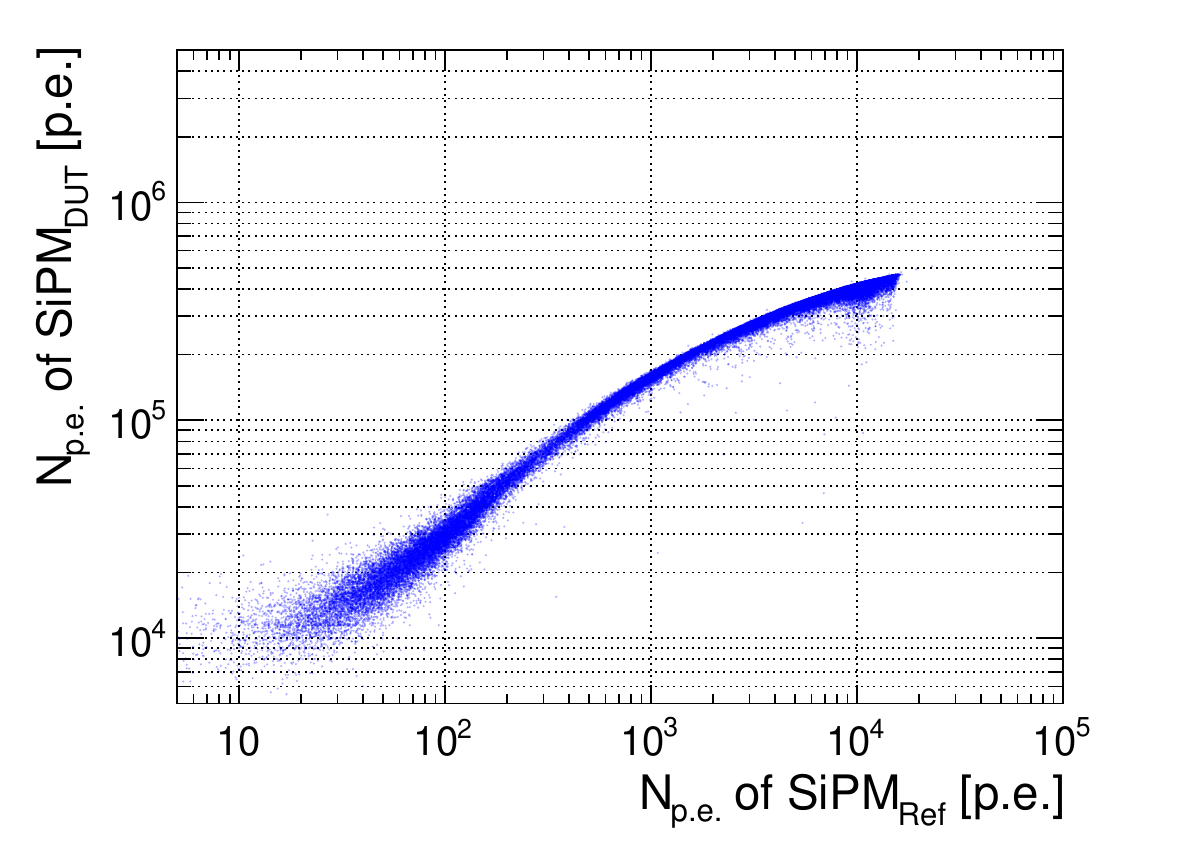}} 
    \subfloat[]{
    \includegraphics[width=0.32\textwidth]{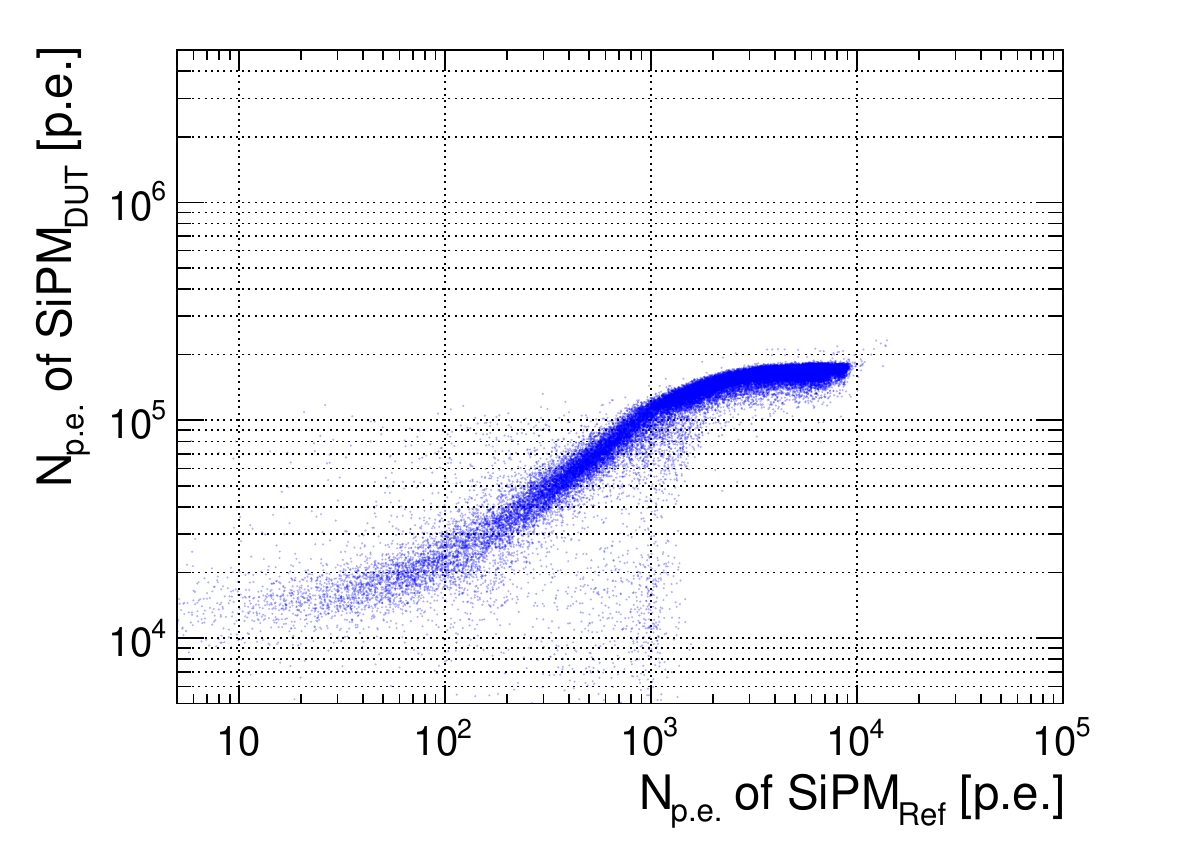}}
    \caption{\label{fig:NPE_raw}~Number of photoelectrons for different unit configurations: (a) S14160-3010PS with $40\times1.5\times1.5~\mathrm{cm}^3$ BGO; (b) S14160-6010PS with $40\times1.5\times1.5~\mathrm{cm}^3$ BGO; (c) S14160-6010PS with $12\times2\times2~\mathrm{cm}^3$ BGO; (d) S14160-6010PS with $12\times2\times2~\mathrm{cm}^3$ BSO; (e) EQR10 with $40\times1.5\times1.5~\mathrm{cm}^3$ BGO; (f) EQR06 with $40\times1.5\times1.5~\mathrm{cm}^3$ BGO.}
\end{figure}

Figure~\ref{fig:NPE_raw} shows the number of photoelectron distributions of the $\mathrm{SiPM_{Ref}}$ and $\mathrm{SiPM_{DUT}}$ signals after the pre-amplifier and SiPM gain calibrations. The horizontal axis represents the $\mathrm{SiPM_{Ref}}$ signal, while the vertical axis corresponds to the $\mathrm{SiPM_{DUT}}$ signal. It can be observed that the $\mathrm{SiPM_{DUT}}$ signal is significantly larger than that of $\mathrm{SiPM_{Ref}}$, which originates from the different light collection efficiencies at the two ends of the crystals. First, an optical filter is installed in front of $\mathrm{SiPM_{Ref}}$, which strongly attenuates the scintillation light. It also increases the distance between SiPM and crystal surface, which reduces the effective acceptance angle of $\mathrm{SiPM_{Ref}}$ and further degrades its light collection efficiency. Second, different optical coupling schemes are employed at the two ends: $\mathrm{SiPM_{DUT}}$ is coupled to the crystal using silicone oil, whereas $\mathrm{SiPM_{Ref}}$ is air-coupled. Owing to its refractive index close to that of the crystal, silicone oil reduces reflection losses at the interface and facilitates photon transmission to SiPM. As a result, $\mathrm{SiPM_{DUT}}$ exhibits a substantially higher light collection efficiency and therefore reaches saturation more readily.

At low $\mathrm{N_{p.e.}}$, the responses of $\mathrm{SiPM_{Ref}}$ and $\mathrm{SiPM_{DUT}}$ follow an approximately linear correlation, indicating that both devices operate in their quasi-linear regions. With increasing signal amplitude, the response of $\mathrm{SiPM_{DUT}}$ starts to deviate from linearity, and the correlation gradually bends towards a smaller slope. The slope in the linear region reflects the ratio of the light collection efficiencies of the two SiPMs. By rescaling the horizontal axis through a linear transformation such that the slope in the linear region is normalized to unity, the corrected $\mathrm{SiPM_{DUT}}$ signal can be used to represent the expected response of $\mathrm{SiPM_{DUT}}$ in the absence of nonlinearity. This procedure allows a direct quantification of the nonlinearity of $\mathrm{SiPM_{DUT}}$ and constitutes the relative light collection efficiency calibration.

\begin{figure}[htbp]
    \centering  
    \subfloat[]{
    \includegraphics[width=0.32\textwidth]{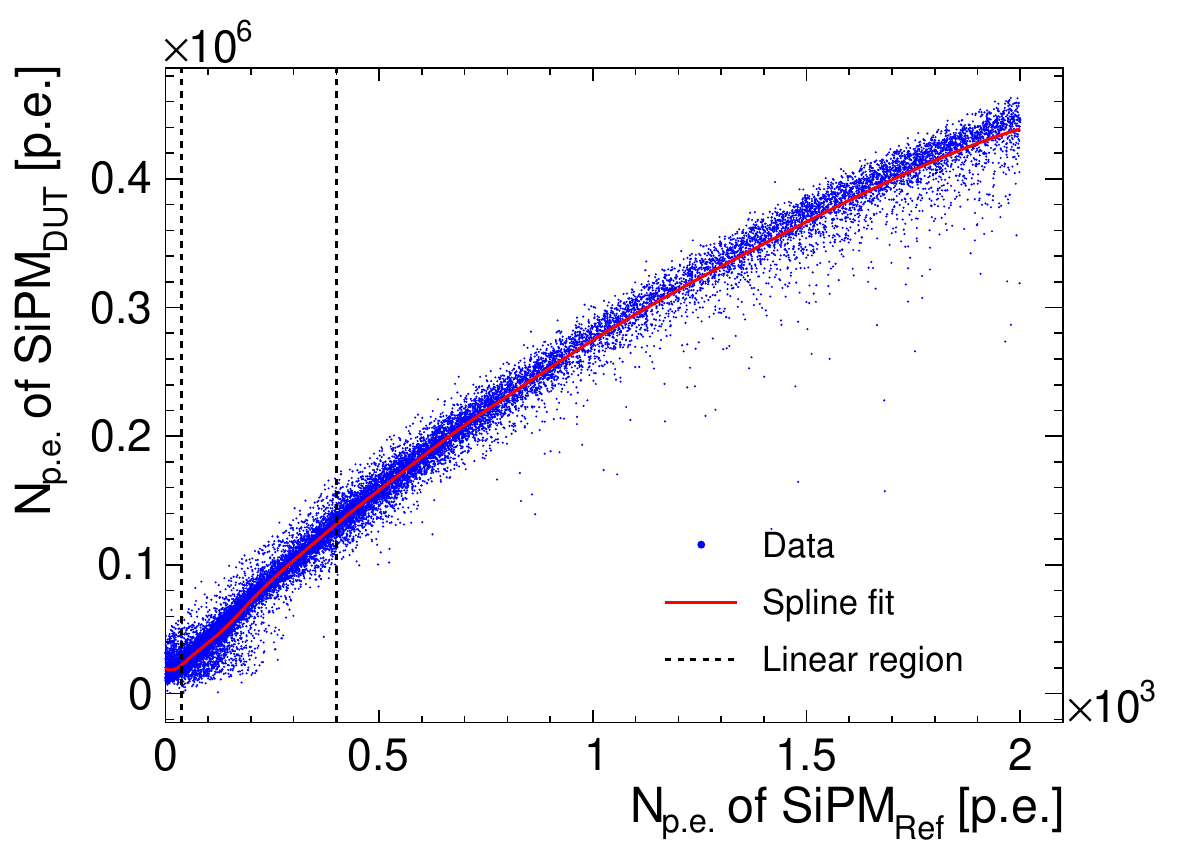}} 
    \subfloat[]{
    \includegraphics[width=0.32\textwidth]{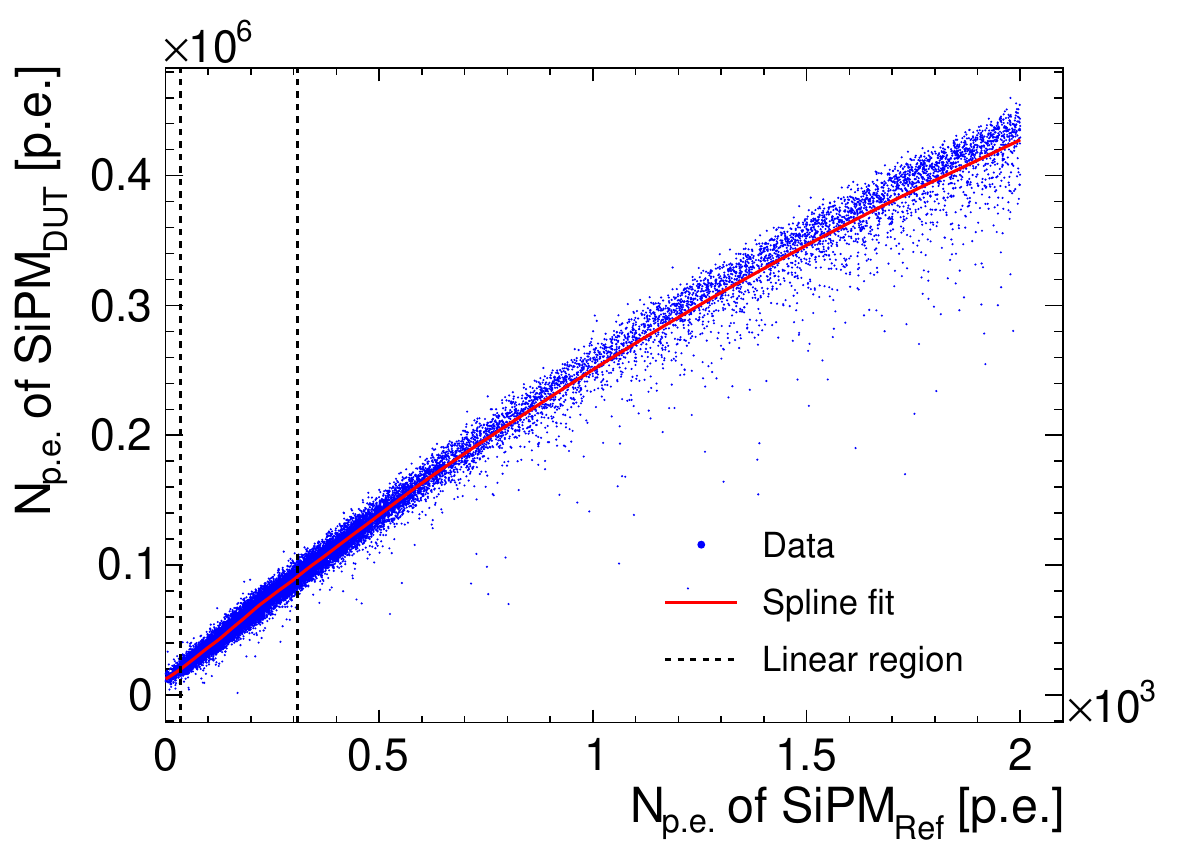}}
    \subfloat[]{
    \includegraphics[width=0.32\textwidth]{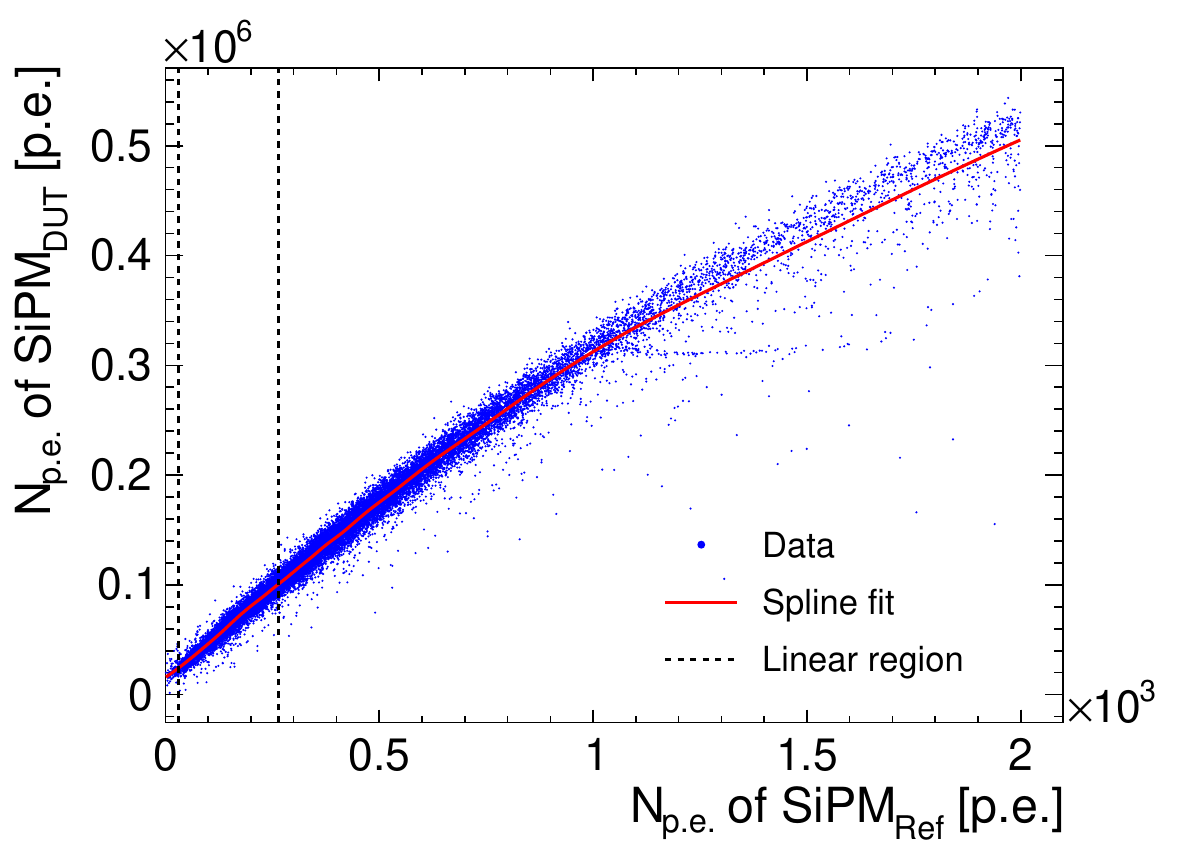}} \\
    \subfloat[]{
    \includegraphics[width=0.32\textwidth]{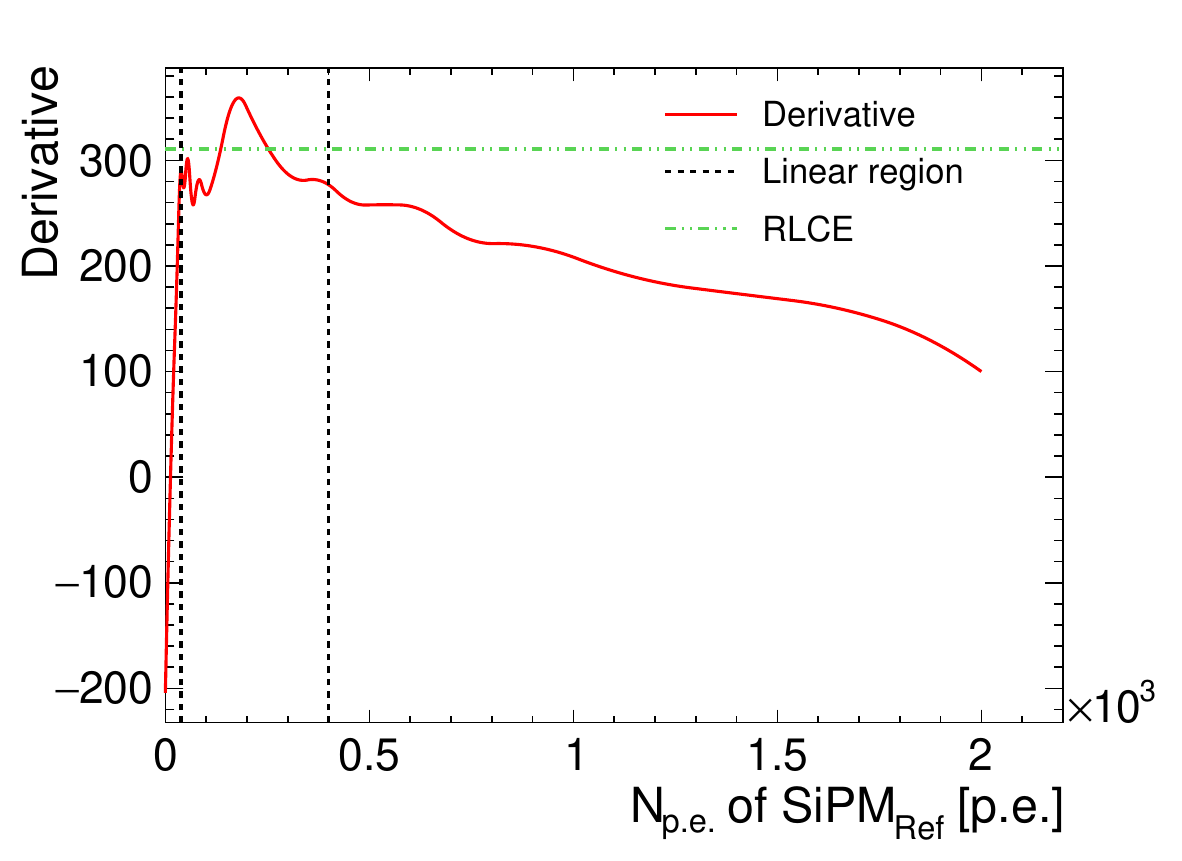}} 
    \subfloat[]{
    \includegraphics[width=0.32\textwidth]{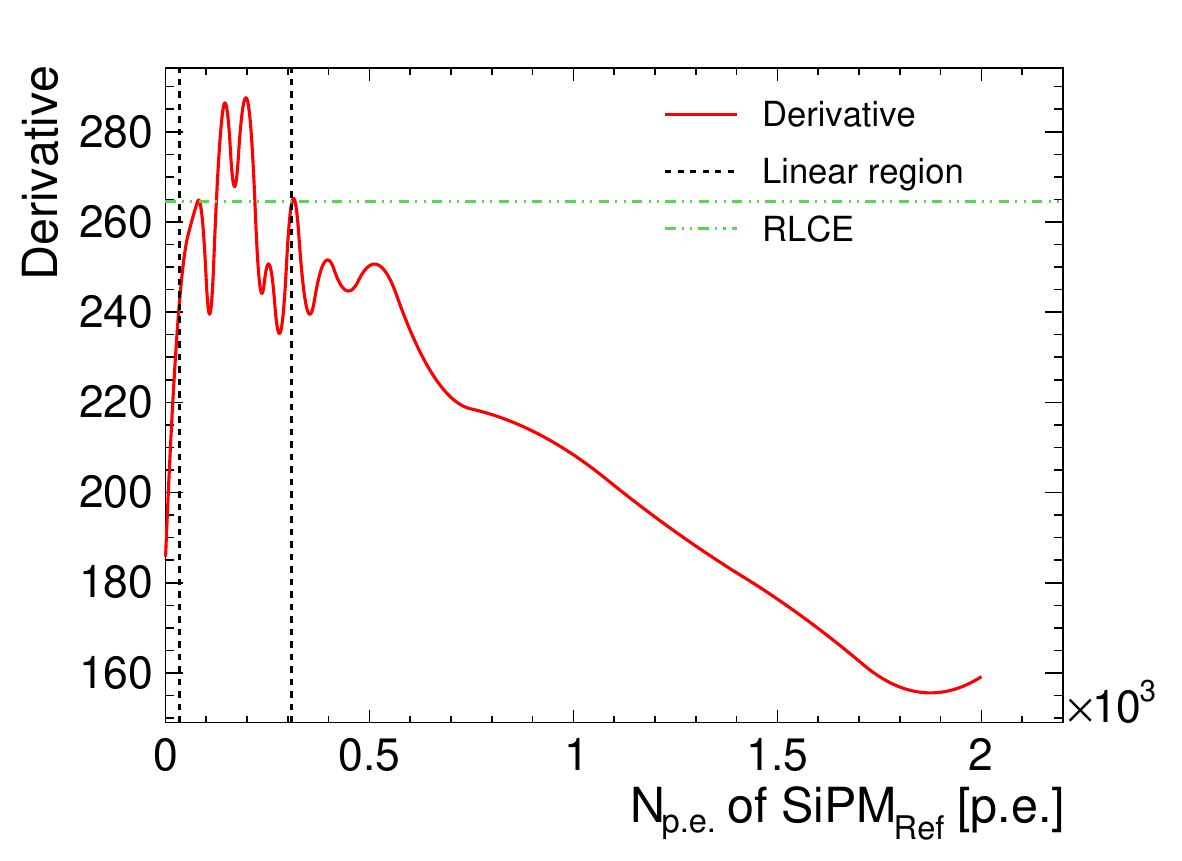}} 
    \subfloat[]{
    \includegraphics[width=0.32\textwidth]{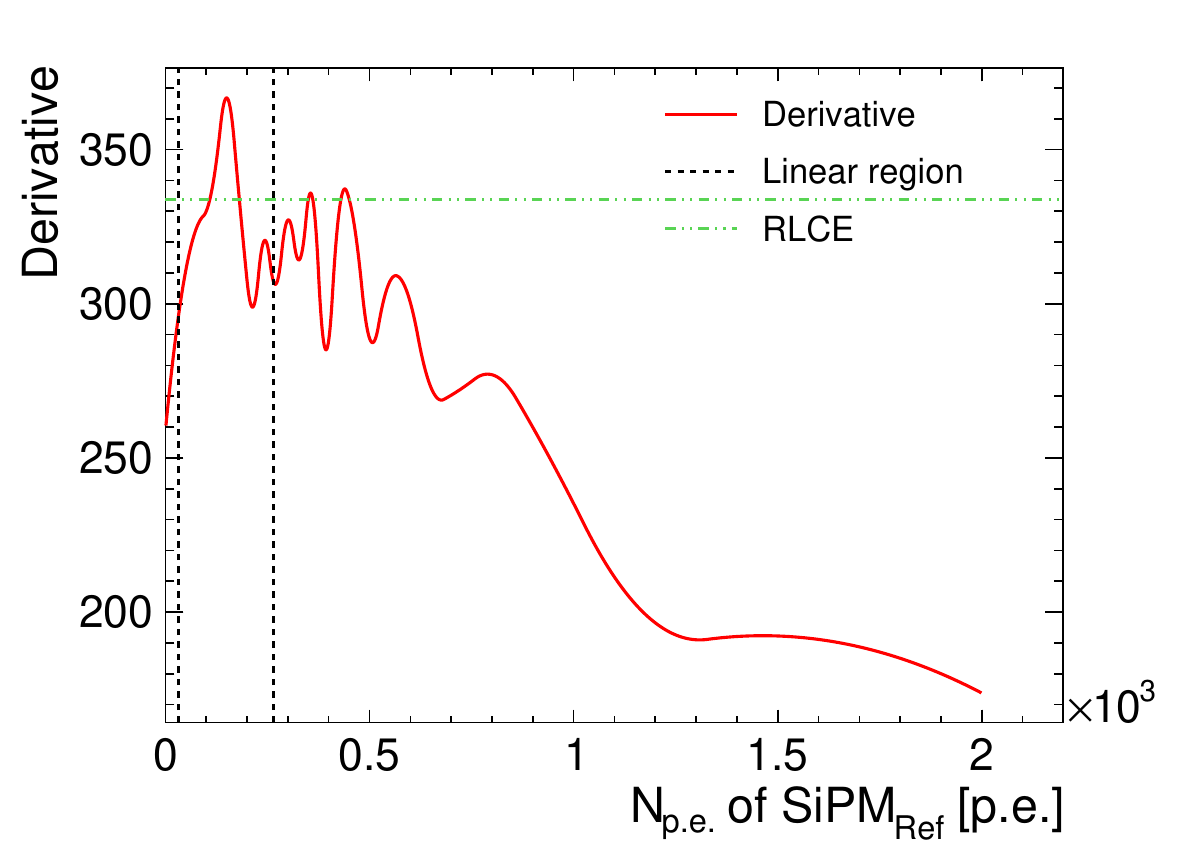}} \\
    \caption{\label{fig:LCE_calibration_a}~Relative light collection efficiency calibration for different crystal--SiPM configurations. The upper panels (a)--(c) show the corrected high-gain $\mathrm{N_{p.e.}}$ distributions together with the spline fits and the identified quasi-linear fitting intervals, while the lower panels (d)--(f) display the corresponding first derivatives of the spline functions and the extracted RLCE values. The configurations correspond to: (a,d) S14160-3010PS with $40\times1.5\times1.5~\mathrm{cm}^3$ BGO; (b,e) S14160-6010PS with $40\times1.5\times1.5~\mathrm{cm}^3$ BGO; (c,f) S14160-6010PS with $12\times2\times2~\mathrm{cm}^3$ BGO.}
\end{figure} 

A crucial step in this procedure is the determination of slope in the quasi-linear region. It should be noted that SiPMs are intrinsically non-linear devices and do not exhibit a perfectly linear operating region. Therefore, a practical criterion is required to define the signal range in which the response can be regarded as approximately linear. In the present analysis, the signal distribution shown in Figure~\ref{fig:NPE_raw} was first fitted using a spline function. The first derivative of the spline with respect to the $\mathrm{SiPM_{Ref}}$ signal was then calculated to evaluate the local slope as a function of signal QDC. Linear fits were subsequently performed in regions where the derivative remained relatively stable. The fitting interval was scanned, and the linear function with a coefficient of determination $R^{2} > 0.995$ was selected as the calibration function. The slope of this function was taken as the relative light collection efficiency.


\begin{figure}[htbp]
    \centering  
    \subfloat[]{
    \includegraphics[width=0.32\textwidth]{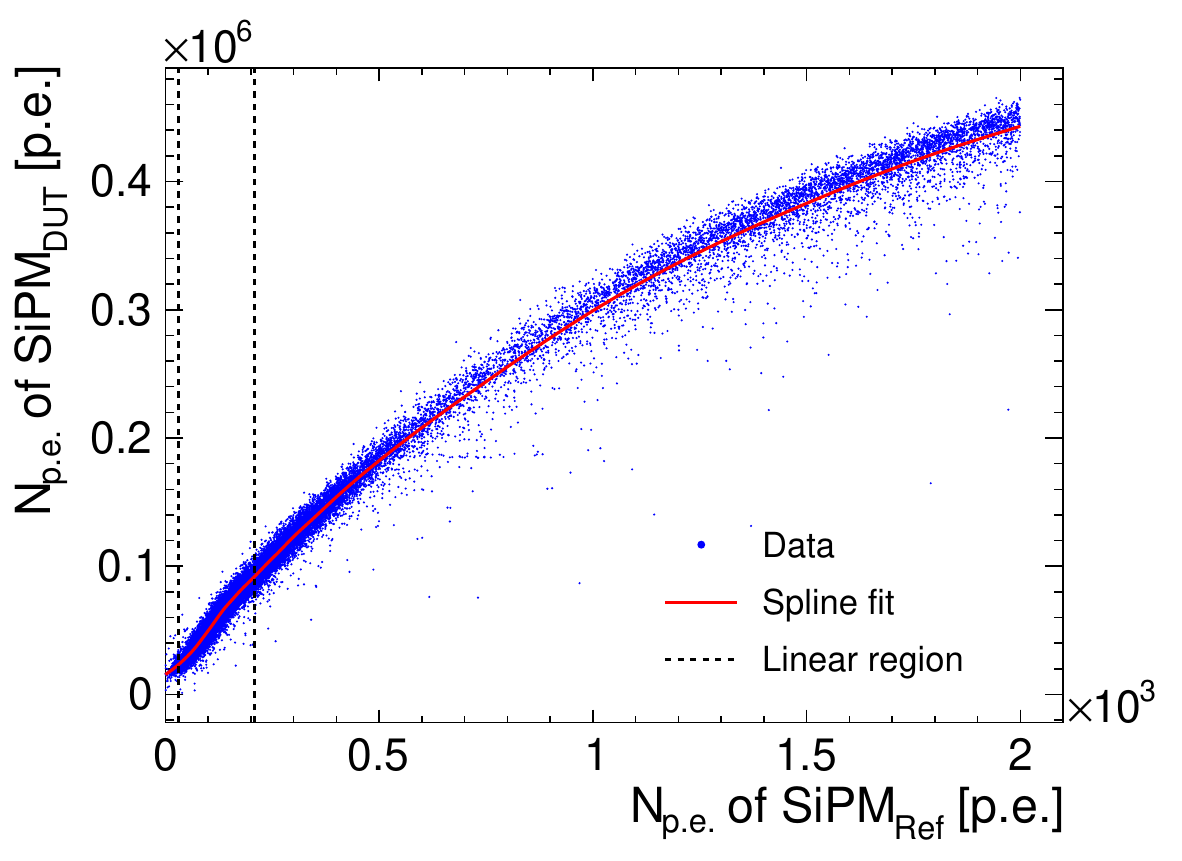}} 
    \subfloat[]{
    \includegraphics[width=0.32\textwidth]{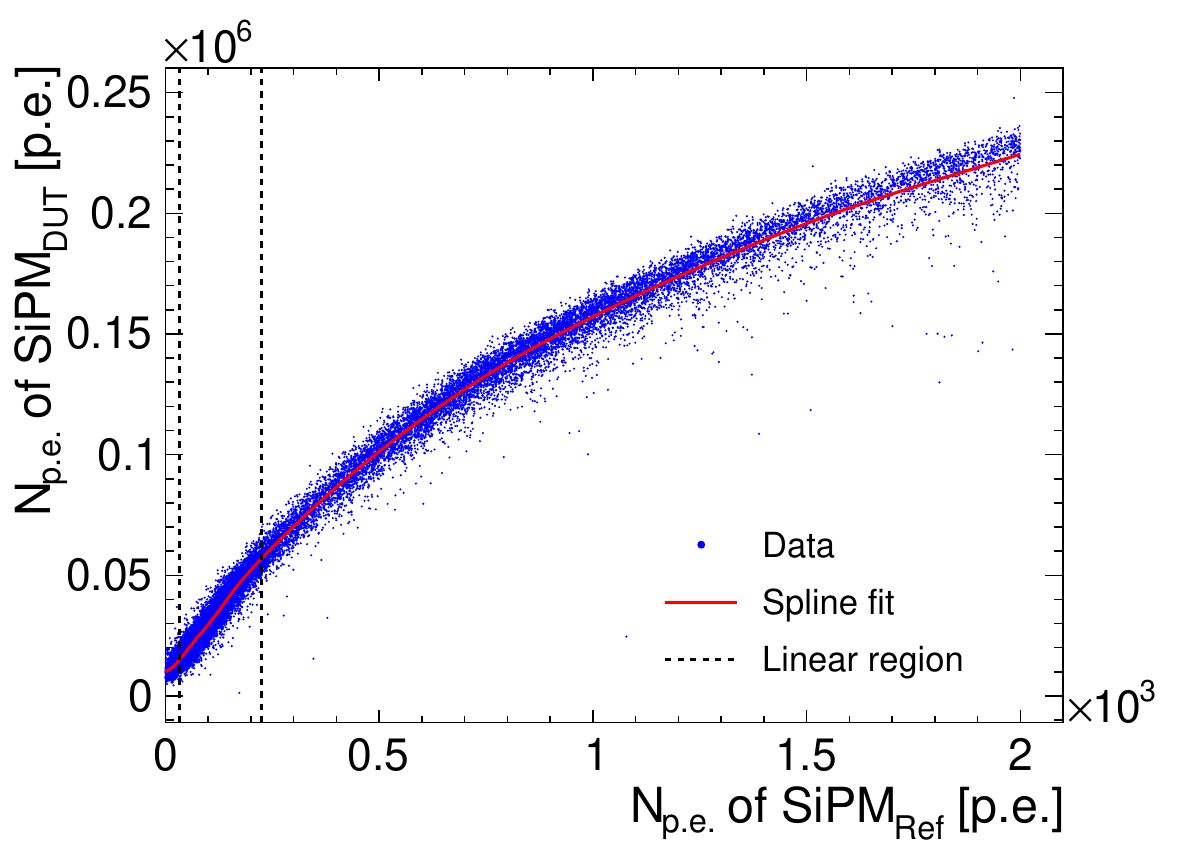}}
    \subfloat[]{
    \includegraphics[width=0.32\textwidth]{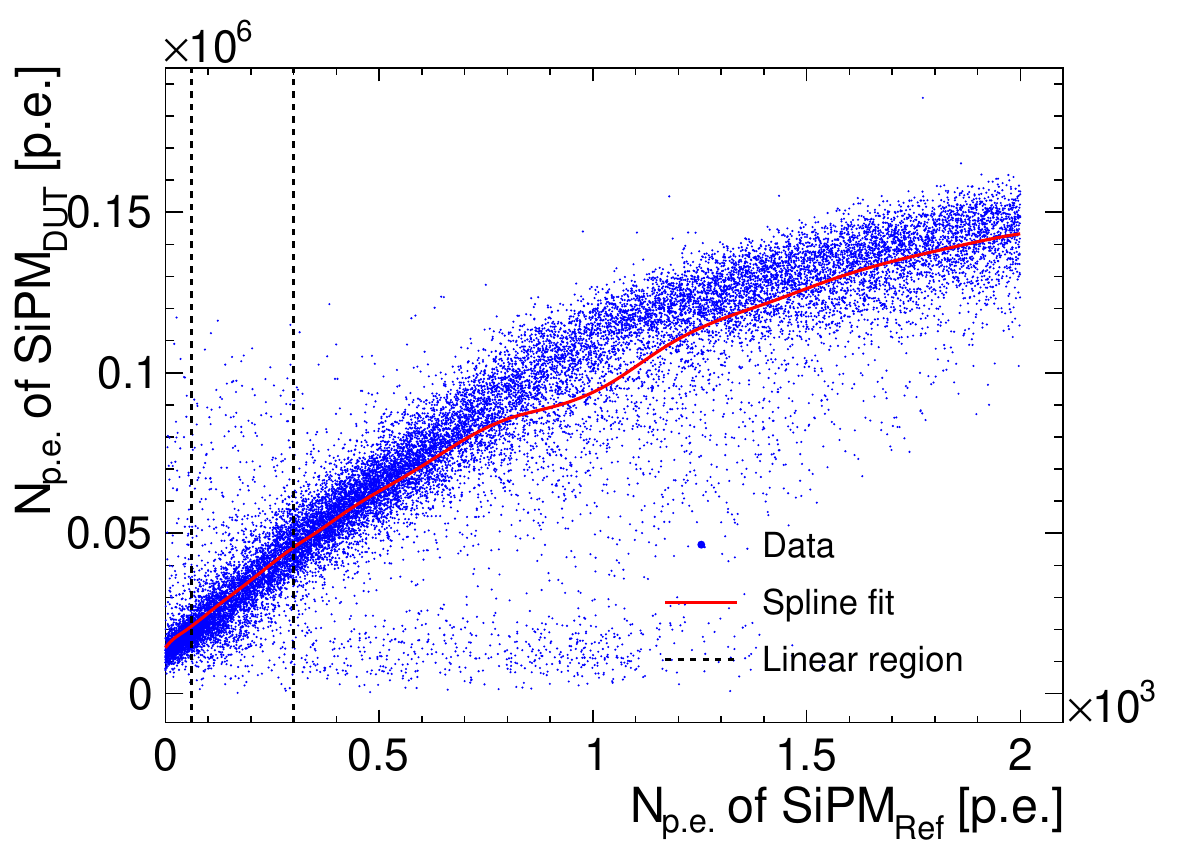}} \\
    \subfloat[]{
    \includegraphics[width=0.32\textwidth]{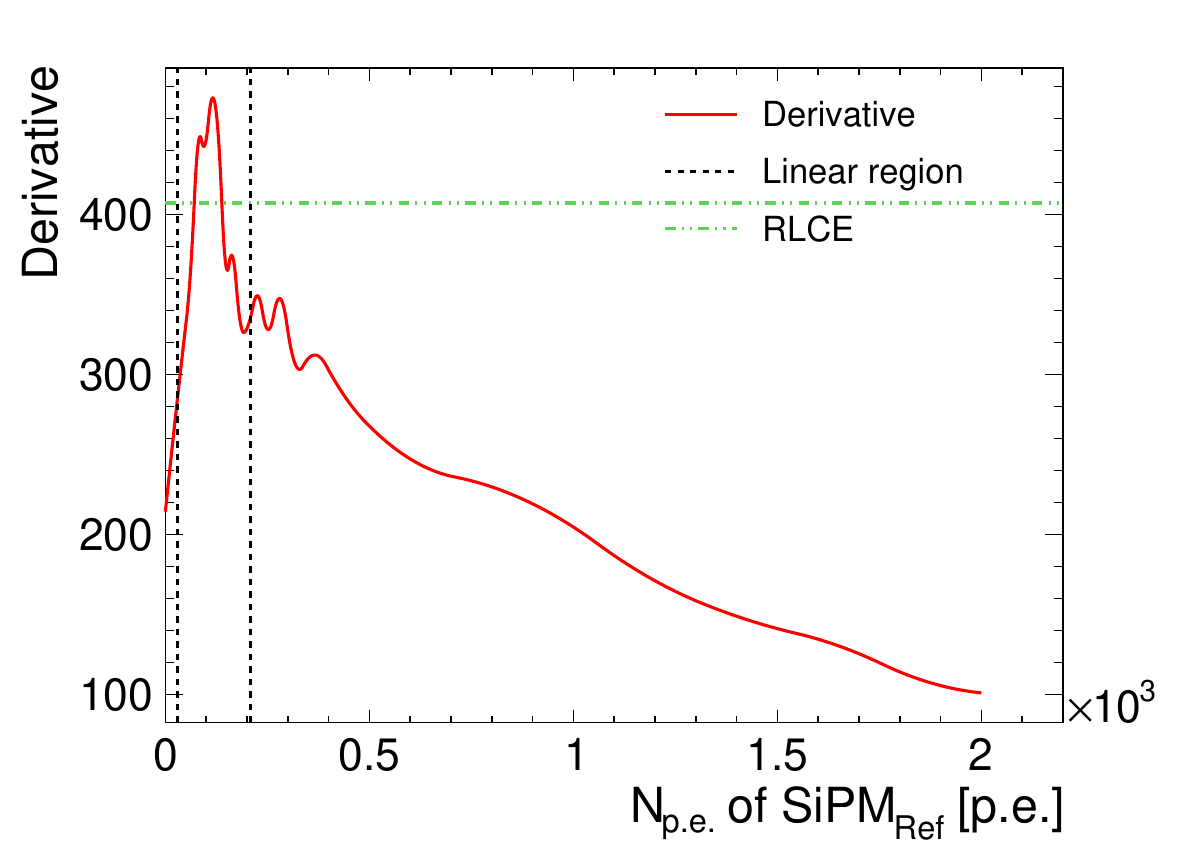}}
    \subfloat[]{
    \includegraphics[width=0.32\textwidth]{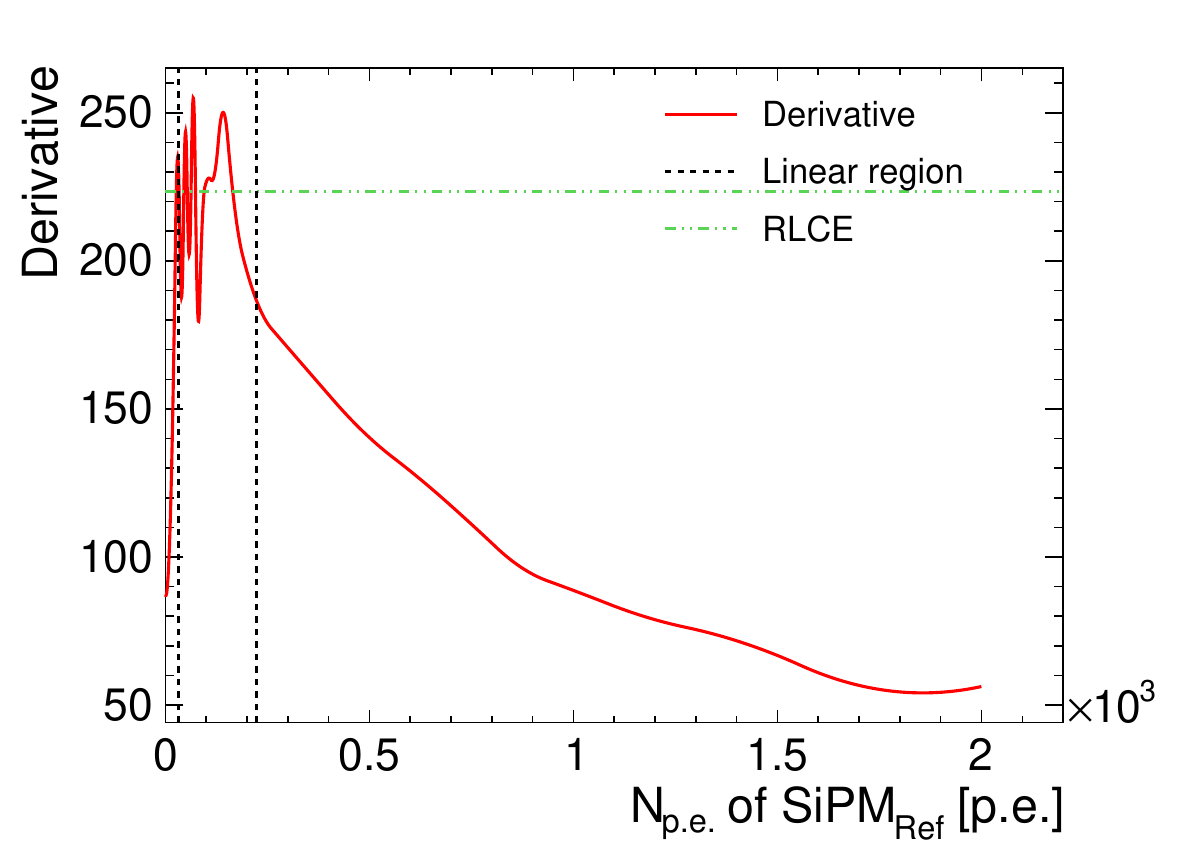}} 
    \subfloat[]{
    \includegraphics[width=0.32\textwidth]{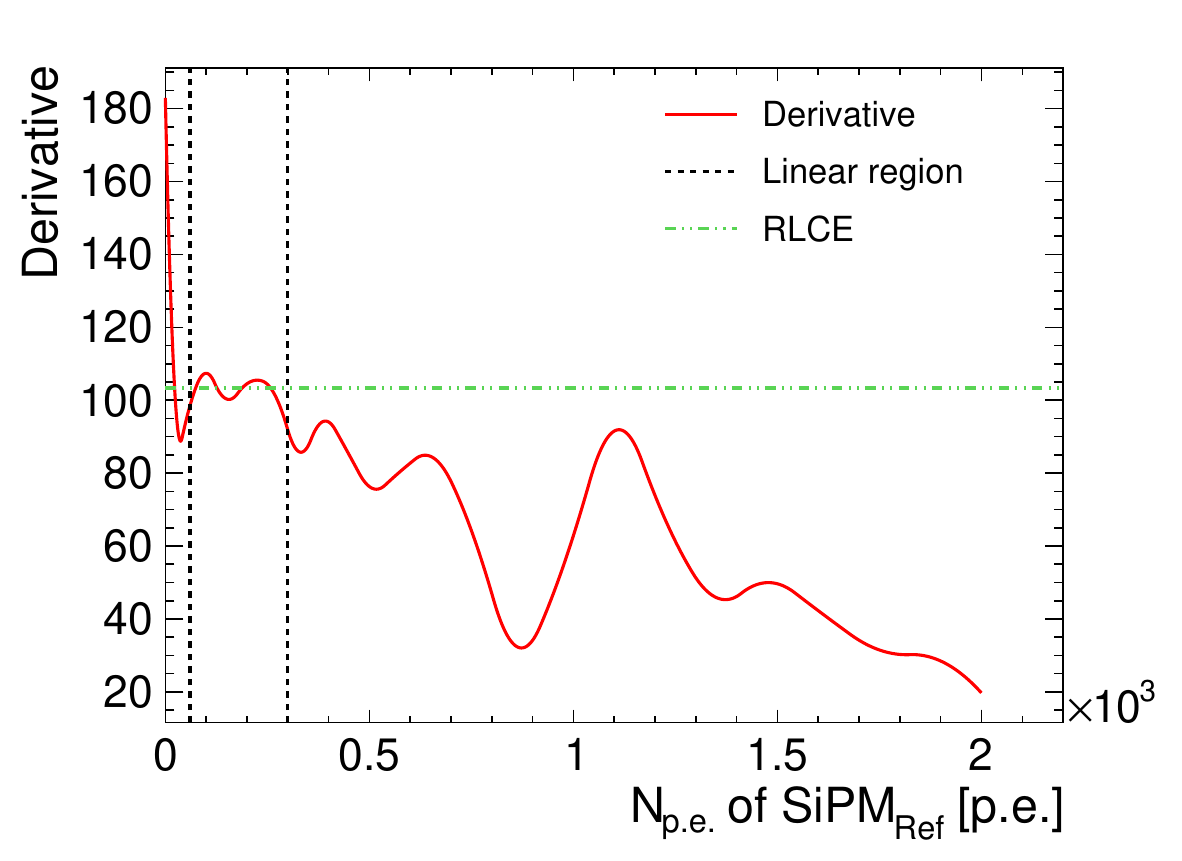}} \\
    \caption{\label{fig:LCE_calibration_b}~Relative light collection efficiency calibration for additional crystal--SiPM configurations. The upper panels (a)--(c) present the corrected high-gain $\mathrm{N_{p.e.}}$ distributions with the corresponding spline fits and quasi-linear regions, and the lower panels (d)--(f) show the first derivatives of the spline functions and the extracted RLCE values. The configurations correspond to: (a,d) S14160-6010PS with $12\times2\times2~\mathrm{cm}^3$ BSO; (b,e) EQR10 with $40\times1.5\times1.5~\mathrm{cm}^3$ BGO; (c,f) EQR06 with $40\times1.5\times1.5~\mathrm{cm}^3$ BGO.}
\end{figure} 

Figure~\ref{fig:LCE_calibration_a} and~\ref{fig:LCE_calibration_b} present the data processing results for different experimental configurations. In each figure, the upper panels (a), (b), and (c) show the $\mathrm{N_{p.e.}}$ distributions together with the corresponding spline fits and the identified quasi-linear fitting intervals. The lower panels display the first derivatives of the spline functions and the extracted relative light collection efficiencies (RLCE) for the respective configurations.

For the Hamamatsu SiPMs (Figure~\ref{fig:LCE_calibration_a} and Figure~\ref{fig:LCE_calibration_b}(a) and (d)), the obtained RLCE values are generally consistent across different crystal configurations, including the 40~cm BGO crystal as well as the 12~cm BGO and BSO crystals, with values ranging from approximately 250 to 450. The observed differences may arise from variations in the coupling conditions. For example, the gaps between different components may vary slightly among configurations, and the performance of the silicone-oil coupling may also differ.

In contrast, the RLCE values obtained with the NDL SiPMs, particularly for the NDL EQR06 11-3030D-S device, are smaller than those of the other configurations. Similar anomalous behaviors were also observed in the subsequent beam-test data analysis and may indicate the presence of device-specific effects.

\section{Beam test results}
\label{sec:beamtest}

The beam test was performed at the CERN SPS H2 beamline. A 300~GeV electron beam was primarily used in order to maximize the energy deposition in the crystals. By rotating the motorized mounting stage, the incident angle of the particles could be varied, allowing the response of the crystal--SiPM detection unit to be studied over a wide range of deposited energies. The incident angle is defined as the angle between the incoming particle direction and the longitudinal axis of the crystal bar. The beam was steered to the geometric center of the crystal, and a tungsten plate with a thickness of 3~mm ($\sim0.86\,X_{0}$) was installed upstream of the crystal (see Figure~\ref{fig:crystal_sipm}). During the beam test, crystals and SiPMs were exchanged as required, while the downstream readout electronics remained unchanged.

\subsection{Energy Absorption}
\label{energy_absorption}



\begin{figure}[htbp]
    \centering  
    \subfloat[]{
    \includegraphics[width=0.32\textwidth]{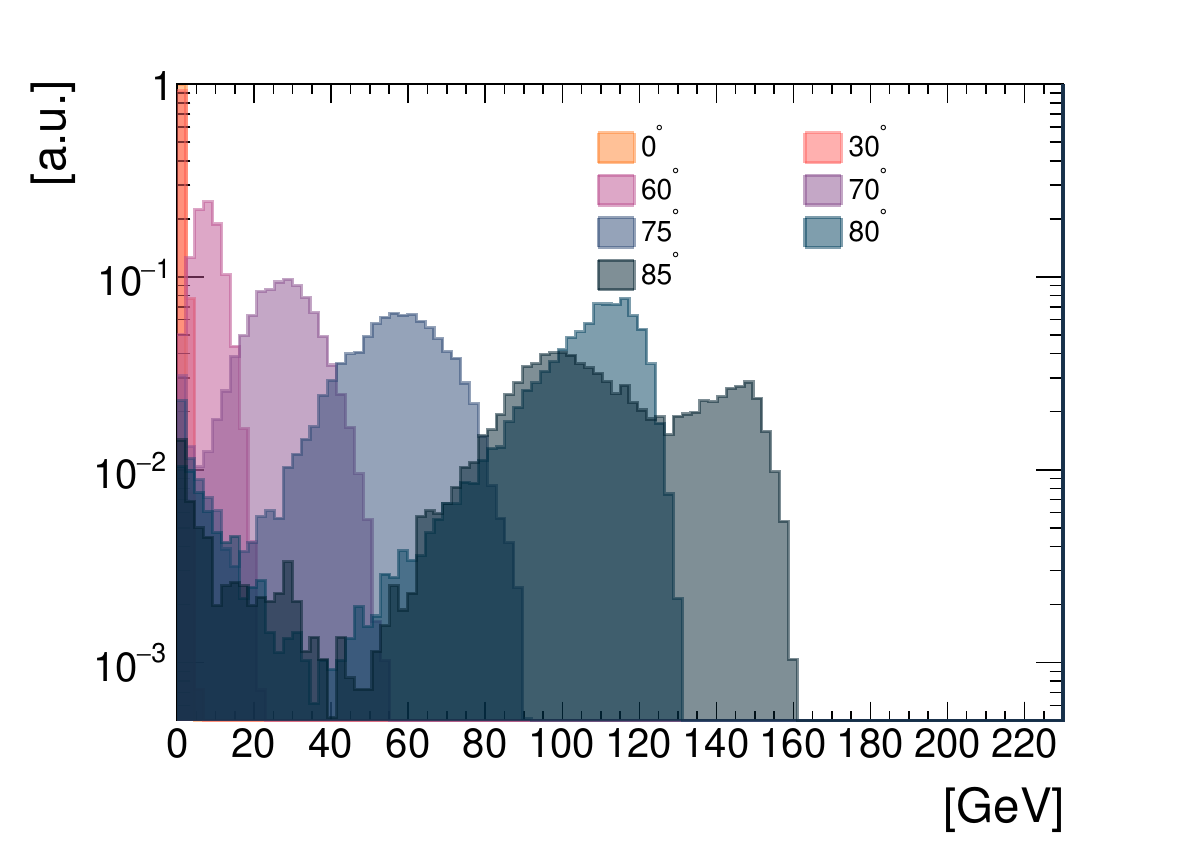}} 
    \subfloat[]{
    \includegraphics[width=0.32\textwidth]{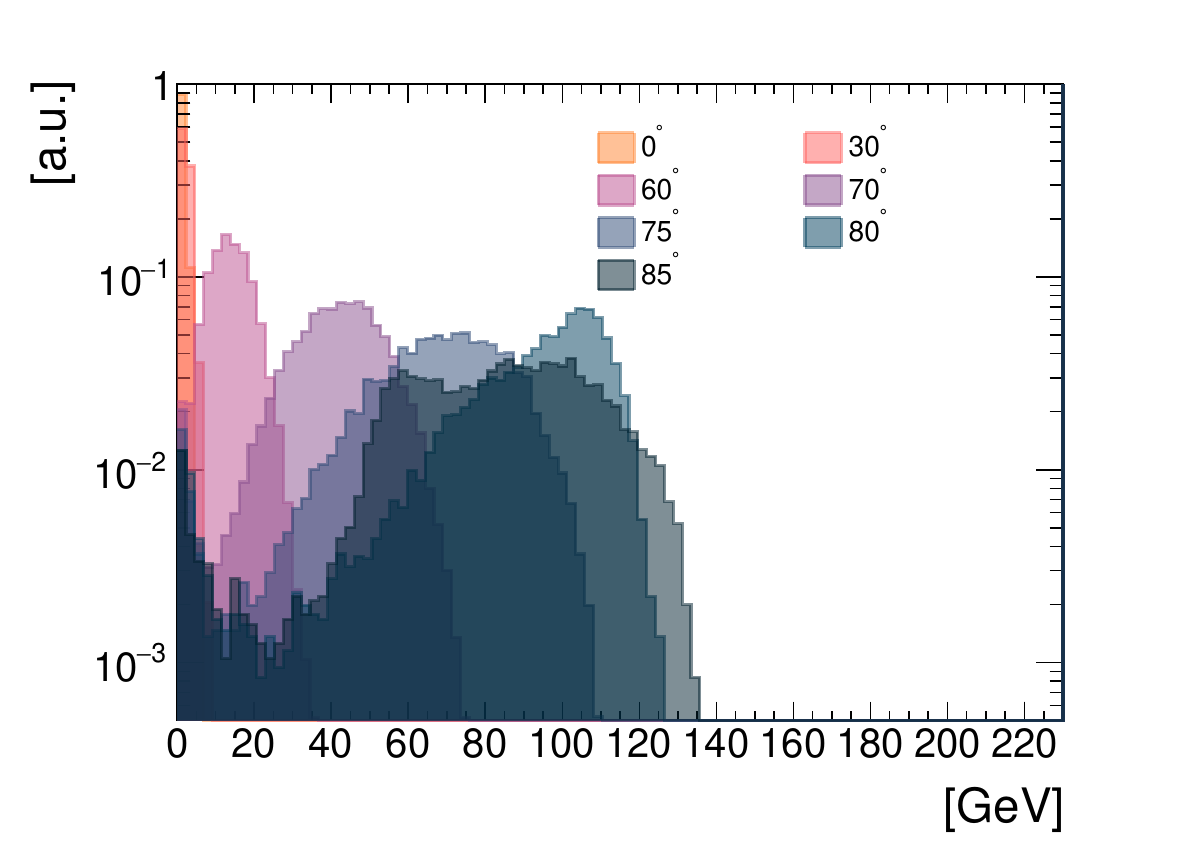}} 
    \subfloat[]{
    \includegraphics[width=0.32\textwidth]{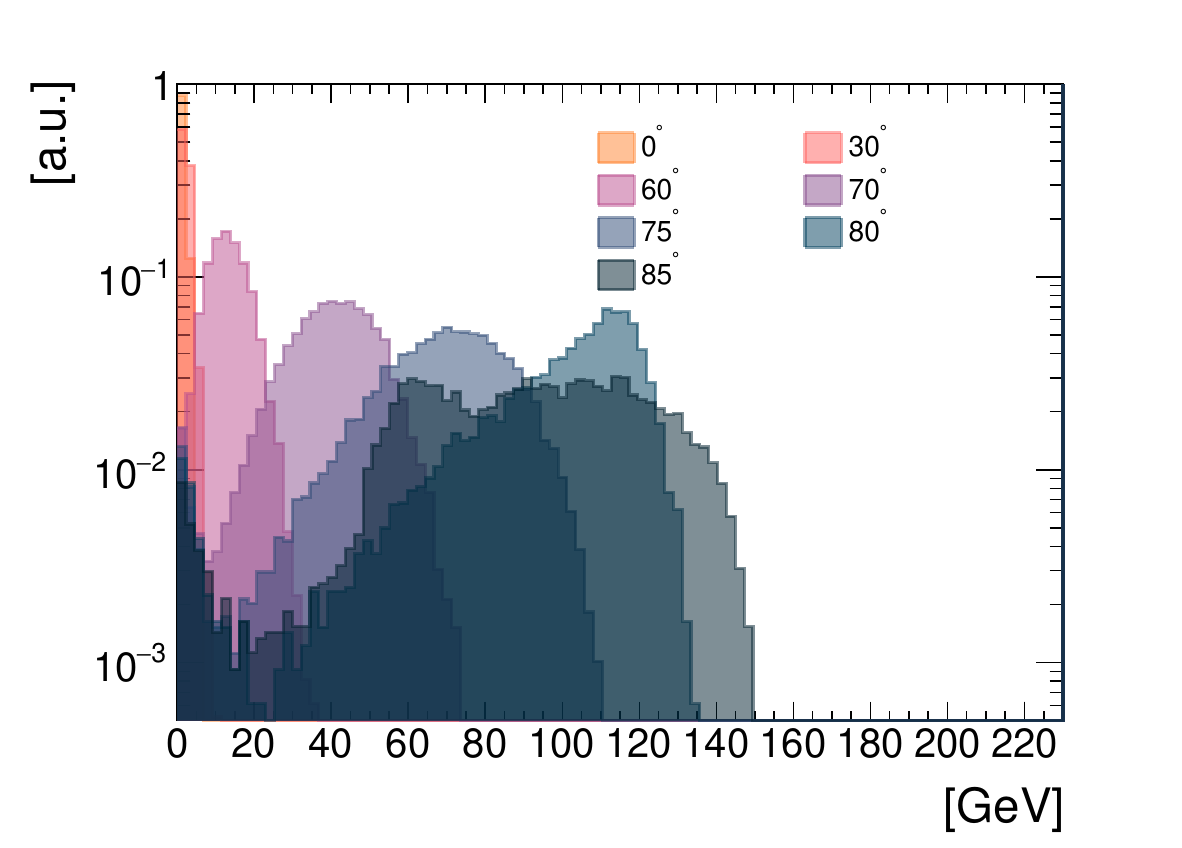}} 
    \caption{\label{fig:energy_angle}~Deposited energy reconstructed from $\mathrm{SiPM_{Ref}}$ at different incident angles for different crystal scintillators coupled to Hamamatsu S14160-6010PS SiPMs.}
\end{figure}

The signal from $\mathrm{SiPM_{Ref}}$, which was attenuated by a low-transmittance optical filter, is expected to remain within its linear response region and therefore provides an accurate measurement of the true energy deposited in the crystal. Using the MIP calibration described in Section~\ref{MIP_calib}, the calibrated $\mathrm{SiPM_{Ref}}$ signals were converted into deposited energy. Figure~\ref{fig:energy_angle} shows the reconstructed energy deposition at different incident angles for different crystal scintillators, all coupled to the same SiPM type (Hamamatsu S14160-6010PS).

For all crystal configurations, the deposited energy increases as the incident angle decreases, reflecting the longer effective path length of the electromagnetic shower within the crystal at shallow angles. The maximum deposited energy can even exceed 100~GeV at sufficiently small incident angles.

\subsection{Number of photoelectrons after calibration}
\label{number_photoelectrons}

\begin{figure}[htbp]
    \centering  
    \subfloat[]{
    \includegraphics[width=0.32\textwidth]{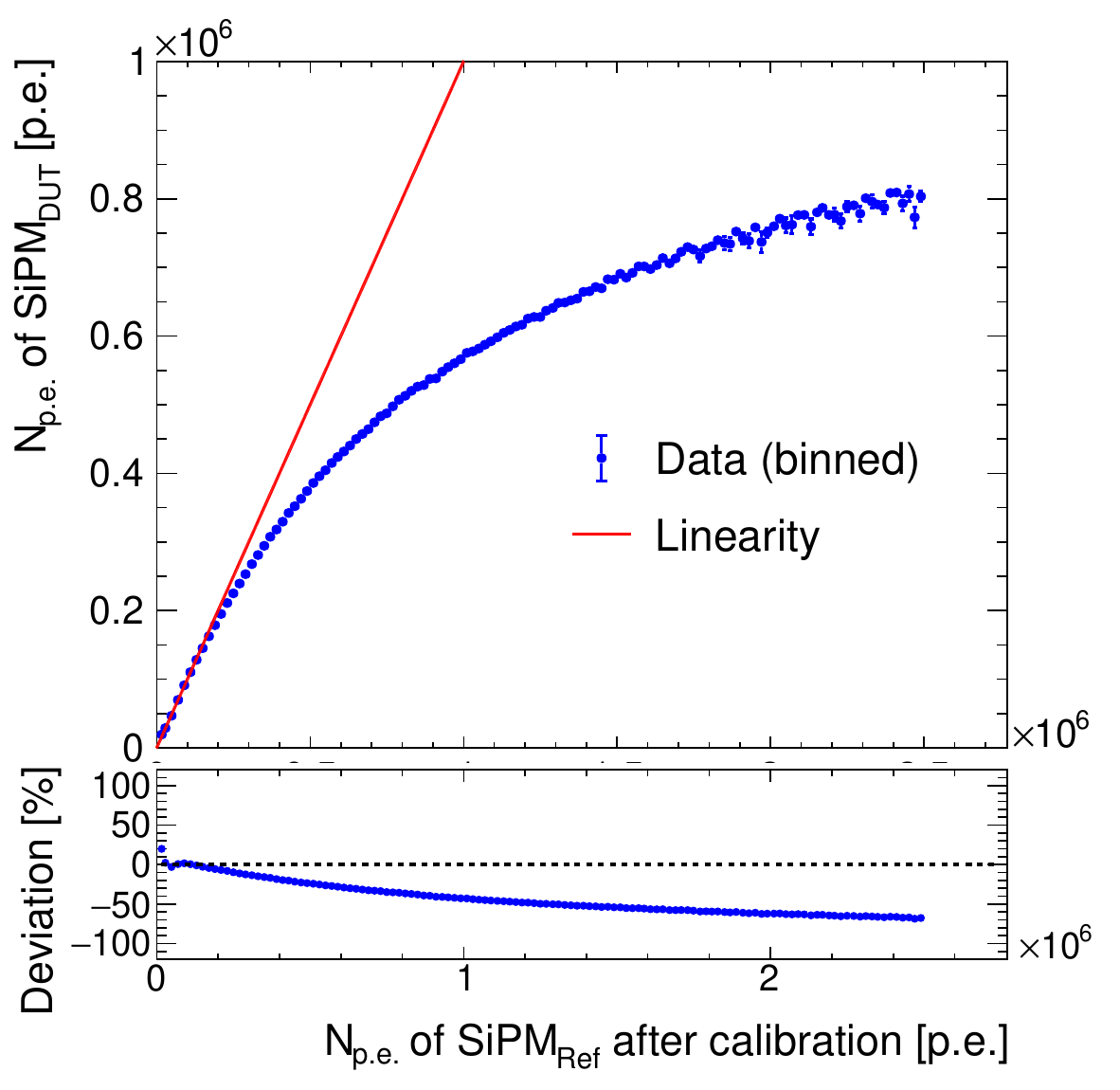}} 
    \subfloat[]{
    \includegraphics[width=0.32\textwidth]{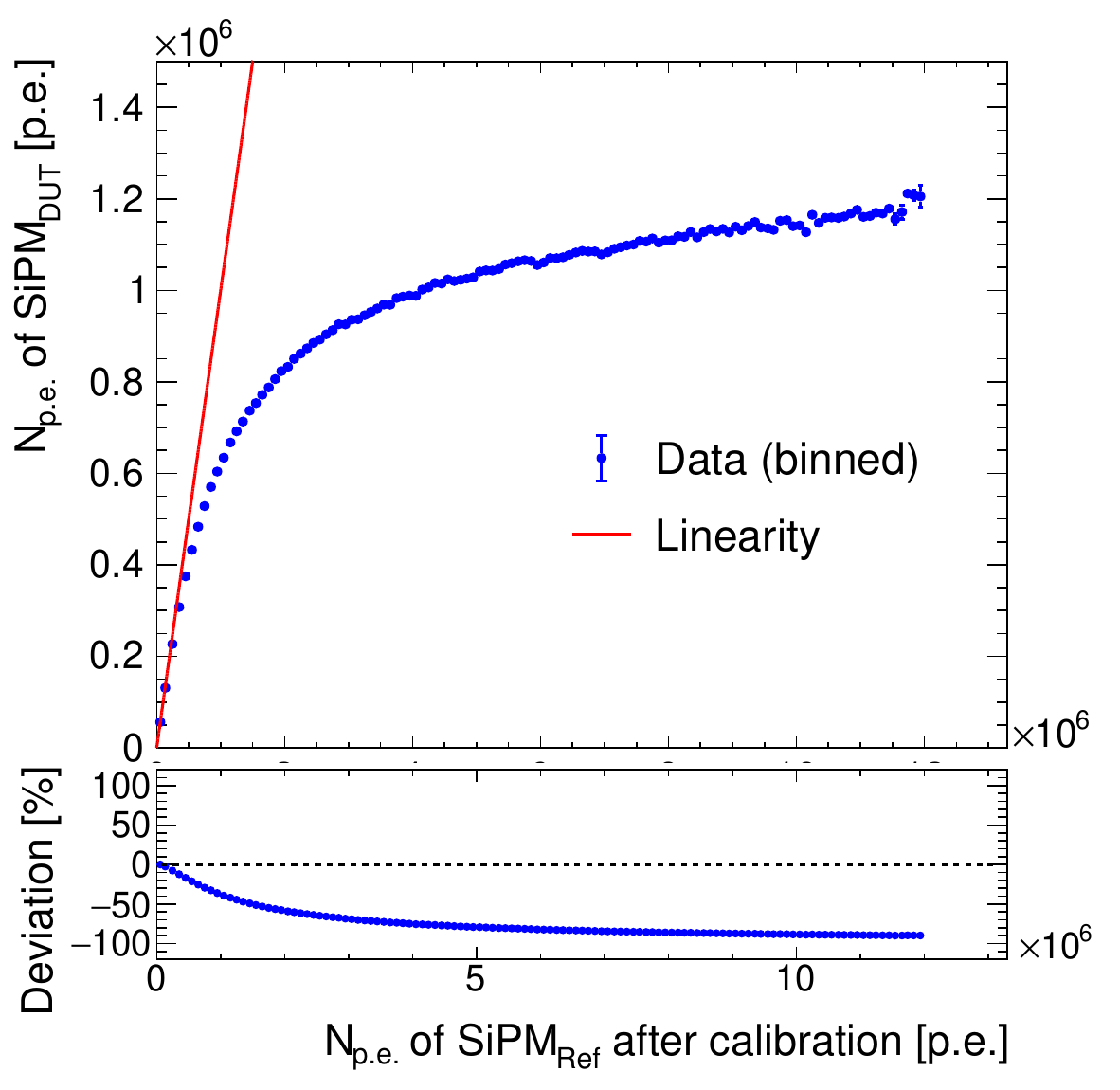}} 
    \subfloat[]{
    \includegraphics[width=0.32\textwidth]{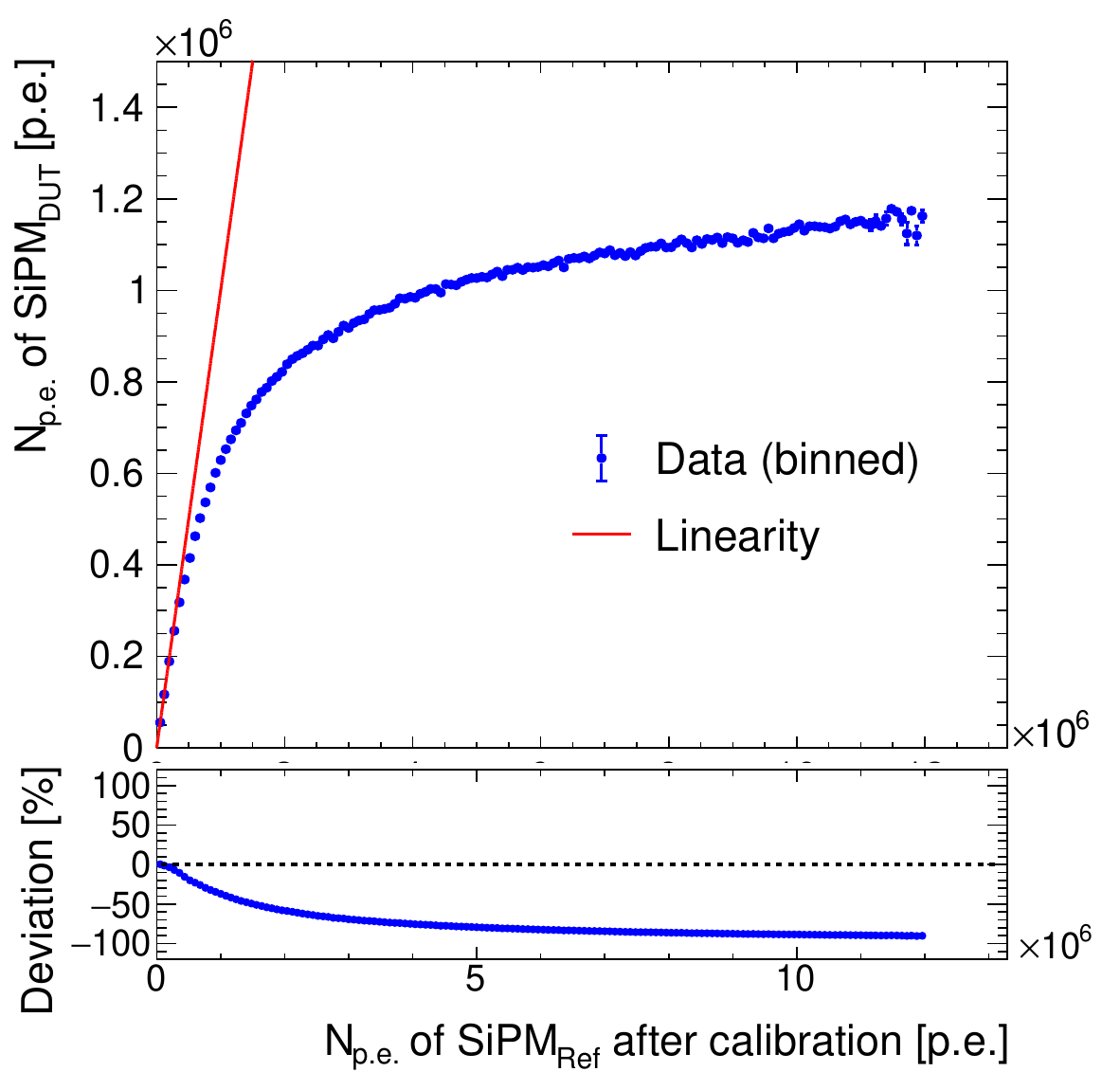}} \\
    \subfloat[]{
    \includegraphics[width=0.32\textwidth]{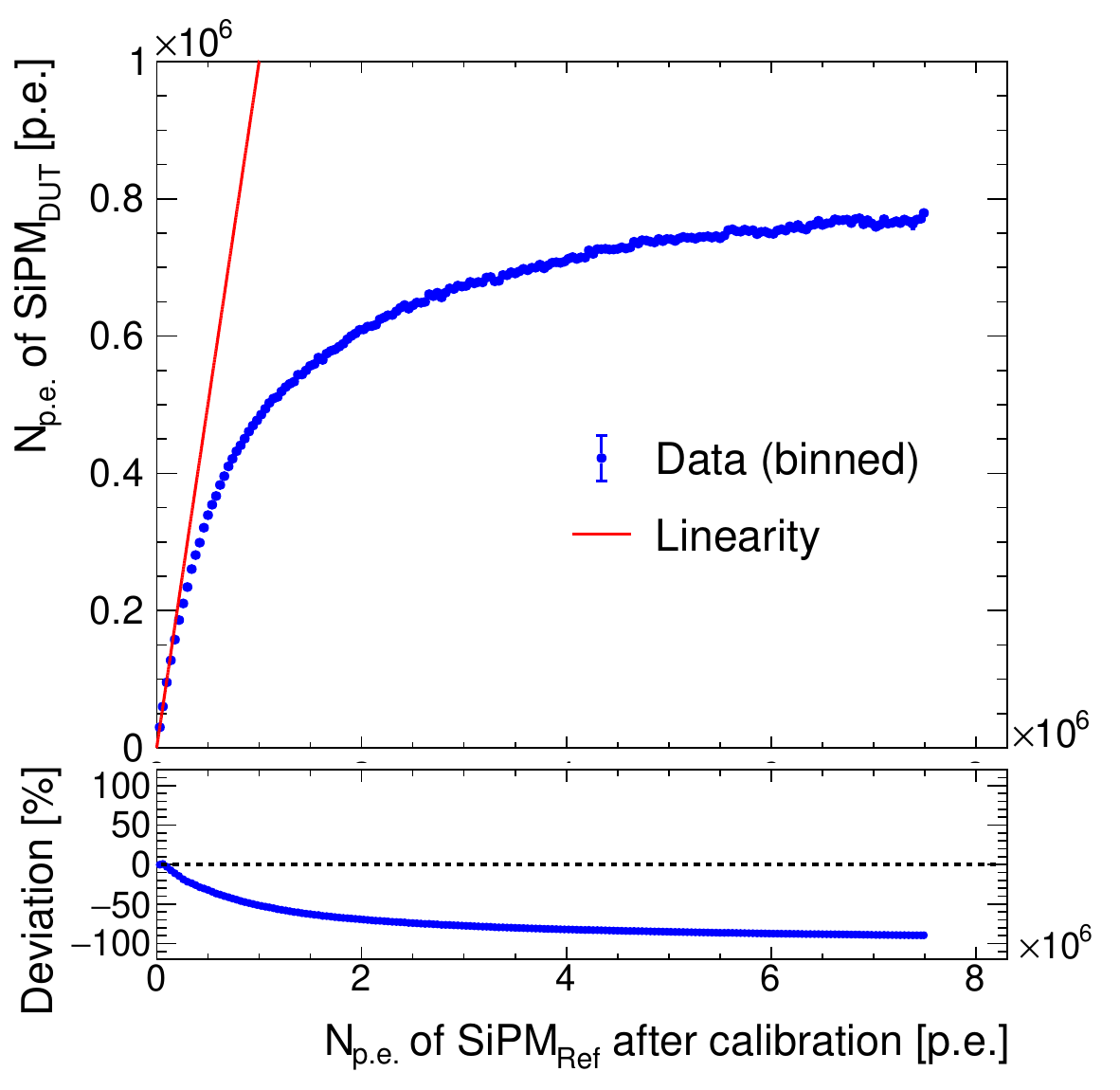}} 
    \subfloat[]{
    \includegraphics[width=0.32\textwidth]{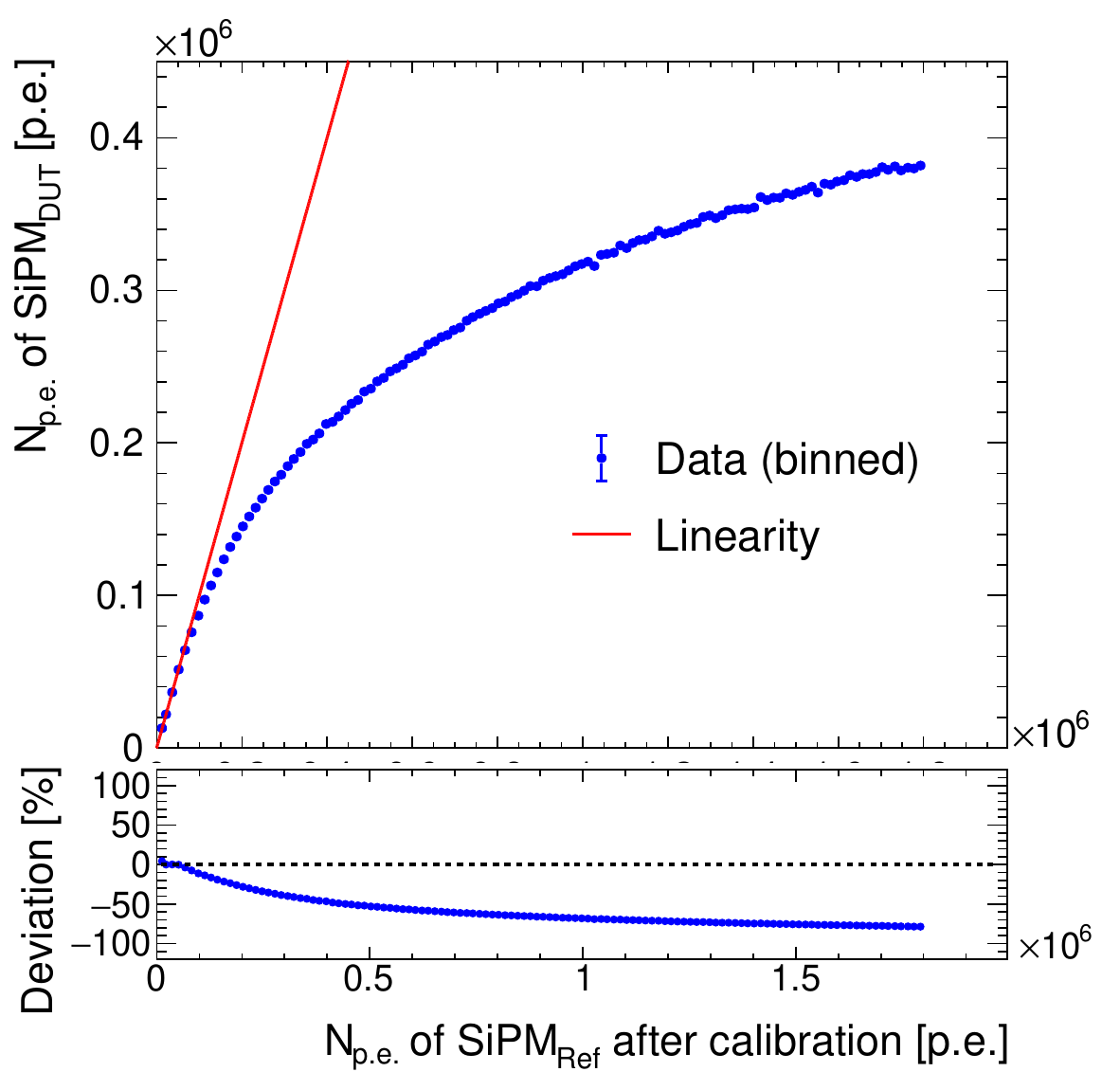}} 
    \subfloat[]{
    \includegraphics[width=0.32\textwidth]{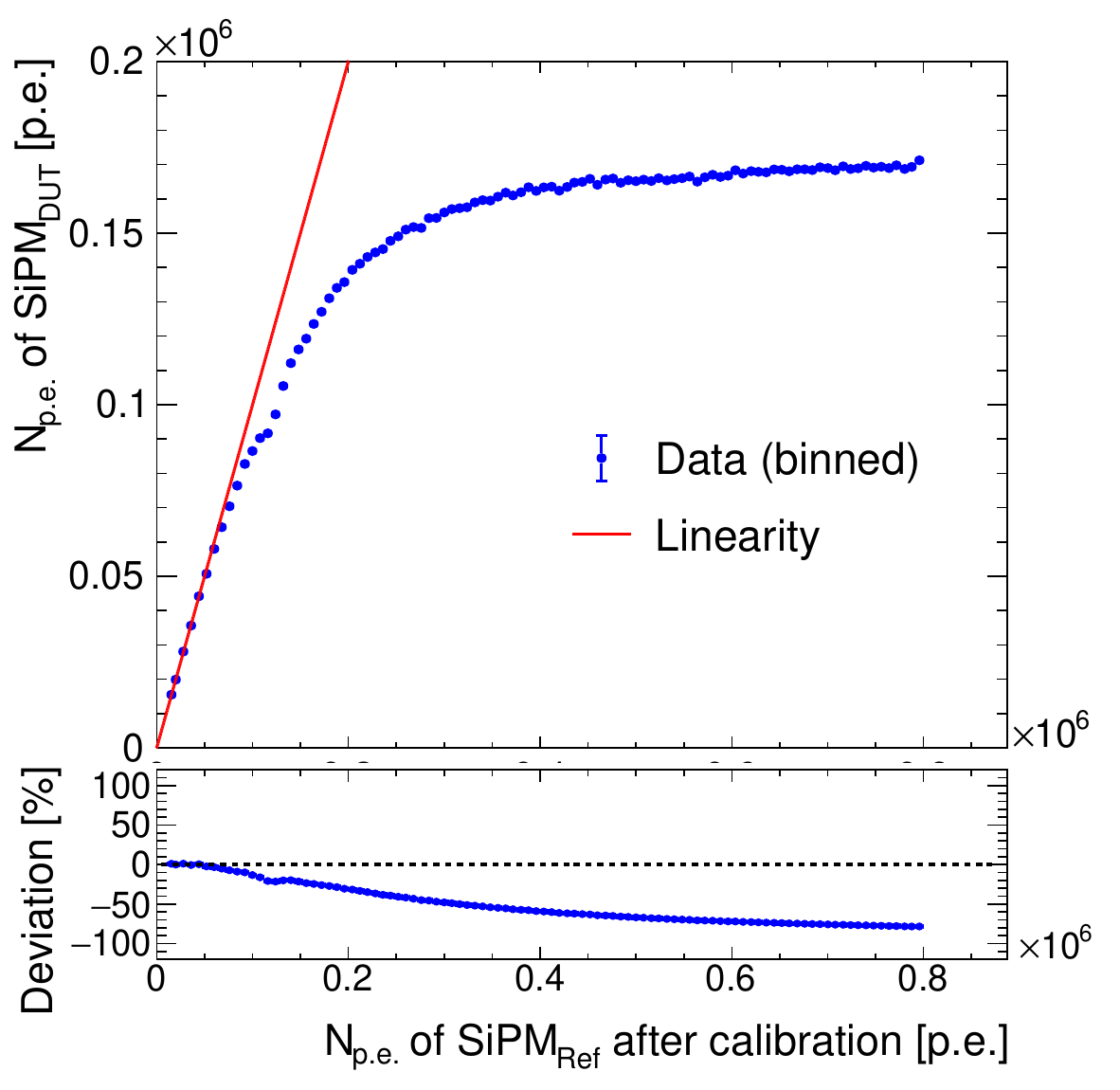}}
    \caption{\label{fig:SiPM_NPE_corrected}~Corrected number of photoelectrons of SiPM responses after linear transformation of the $\mathrm{SiPM_{Ref}}$ signal: (a) S14160-3010PS with $40\times1.5\times1.5~\mathrm{cm}^3$ BGO; (b) S14160-6010PS with $40\times1.5\times1.5~\mathrm{cm}^3$ BGO; (c) S14160-6010PS with $12\times2\times2~\mathrm{cm}^3$ BGO; (d) S14160-6010PS with $12\times2\times2~\mathrm{cm}^3$ BSO; (e) EQR10 with $40\times1.5\times1.5~\mathrm{cm}^3$ BGO; (f) EQR06 with $40\times1.5\times1.5~\mathrm{cm}^3$ BGO.}
\end{figure}

Using the calibration results described in Section~\ref{sec:calibration}, the corrected SiPM responses are presented in Figure~\ref{fig:SiPM_NPE_corrected}, showing the measurement results for different crystal-SiPM combinations. In these figures, the upper panels display the relationship between the responses of SiPM$_{\text{DUT}}$ and SiPM$_{\text{Ref}}$, while the lower panels show the response deviation of SiPM$_{\text{DUT}}$, which can be used to quantify the response linearity of SiPM$_{\text{DUT}}$ at different photoelectron intensities. Table~\ref{tab:nonlinearity_summary} summarizes the linearity deviations of SiPM$_{\text{DUT}}$ at $5\times10^{5}$ photoelectrons for the six different combinations shown in Figure~\ref{fig:SiPM_NPE_corrected}.

\begin{table}[htbp]
\centering
\fontsize{9}{13}\selectfont
\caption{\label{tab:nonlinearity_summary}~Linearity deviations of SiPM$_{\text{DUT}}$ at $5\times10^{5}$ photoelectrons for different crystal-SiPM combinations.}
    \begin{tabular}{ccc}
        \toprule
        \makecell[c]{Crystal} & \makecell[c]{SiPM} & \makecell[c]{Deviation at $5\times10^{5}$ p.e.}\\ 
        \midrule
        $40\times1.5\times1.5~\mathrm{cm}^3$ BGO & HPK S14160-3010PS & -24.0\% \\
        $40\times1.5\times1.5~\mathrm{cm}^3$ BGO & HPK S14160-6010PS & -19.1\% \\
        $12\times2\times2~\mathrm{cm}^3$ BGO & HPK S14160-6010PS & -19.0\% \\
        $12\times2\times2~\mathrm{cm}^3$ BSO & HPK S14160-6010PS & -32.2\% \\
        $40\times1.5\times1.5~\mathrm{cm}^3$ BGO & NDL EQR10 11-3030D-S & -66.9\% \\
        $40\times1.5\times1.5~\mathrm{cm}^3$ BGO & NDL EQR06 11-3030D-S & -53.0\% \\
        \bottomrule
    \end{tabular}
\end{table}


For the two Hamamatsu SiPMs coupled to BGO crystals (Figure~\ref{fig:SiPM_NPE_corrected} (a), (b), (c)) with the same pixel density, the deviations from linearity at $5\times10^{5}$ photoelectrons are both ranging from -24\% to -19\%, which are close to each other. The two Hamamatsu SiPM models used in this study have the same pixel density. For a given number of photoelectrons, the 6~mm SiPM(S14160-6010PS) experiences a lower photon density compared to the 3~mm SiPM(S14160-3010PS), which tends to mitigate saturation effects. On the other hand, the larger junction capacitance of the 6~mm SiPM results in wider single-photon signal pulses and a longer pixel recovery time, making it more prone to saturation. The combined effects of these two factors lead to similar saturation behavior for the two devices. In contrast, for the two NDL SiPMs, the nonlinearity at $5\times10^{5}$ photoelectrons exceeds approximately 50\%. Even with identical pixel densities, the NDL SiPMs should nominally exhibit better linearity due to their smaller junction capacitance and shorter pixel recovery time, not to mention the NDL EQR06 SiPM with its 6~$\mu$m pixel pitch. The underlying reason for this observed behavior is not yet fully understood; however, it is consistent with trends observed in our previous laser-based measurements~\cite{Zhao:2024ohy}.

A comparison of measurements performed with the same SiPM type (Hamamatsu S14160-6010PS) but different crystal scintillators (Figure~\ref{fig:SiPM_NPE_corrected} (b) and~(d)) shows that the BSO-coupled configuration exhibits stronger nonlinearity than the BGO case. This effect is primarily attributed to the faster scintillation decay of BSO, which leads to a higher instantaneous pixel occupancy. Furthermore, the results obtained with 12~cm and 40~cm BGO crystals are very similar, indicating that SiPM nonlinearity is dominated by the intrinsic scintillation decay time, while the impact of photon transport time differences associated with the crystal length is comparatively small.

\subsection{Systematic uncertainties}
\label{systematic_uncertainties}

The data points shown in Figure~\ref{fig:SiPM_NPE_corrected} quantify the non-linear response of the SiPMs to a given scintillation light output. One important potential application of these measurements is the correction of SiPM nonlinearity in practical scintillator--SiPM detector systems, with the aim of improving the accuracy of energy reconstruction. In the present analysis, a sequence of offline calibrations was applied to convert the measured signals into photoelectron counts. Each step in this calibration chain contributes to the overall uncertainty of the derived SiPM nonlinearity.

Systematic uncertainties arising from the individual calibration procedures, as well as from intrinsic properties of the crystals, were evaluated and are summarized in Table~\ref{tab:sys_error}. The listed contributions include uncertainties associated with the pre-amplifier calibration, the SiPM gain calibration, and the non-uniform response of the crystal. For the pre-amplifier and SiPM calibrations, the quoted uncertainties correspond to the standard deviation of the relative residual distributions obtained from the respective calibration fits. The crystal non-uniformity accounts for variations in detector response when particles traverse different regions of the crystal, predominantly along its longitudinal axis. This effect depends on both the crystal growth and processing, as well as on the transport and collection of scintillation light within the crystal volume.

For each experimental configuration. The overall uncertainty is found to be within 3\%, which is combined with all the individual contributions in Table~\ref{tab:sys_error}. 

\begin{table}[htbp]
\centering
\fontsize{9}{13}\selectfont
\caption{\label{tab:sys_error}~Summary of systematic uncertainty contributions.}
    \begin{tabular}{lc}
        \toprule
        \makecell[c]{Source} & \makecell[c]{Uncertainty}\\ 
        \midrule
        Pre-amplifier calibration                                       & 1.5\% \\
        \midrule
        SiPM calibration -- S14160-3010PS                               & 0.8\% \\
        SiPM calibration -- S14160-6010PS                               & 1.4\% \\
        SiPM calibration -- EQR06 11-3030D-S                            & 1.7\% \\
        SiPM calibration -- EQR10 11-3030D-S                            & 1.3\% \\
        \midrule
        Crystal non-uniformity -- $40\times1.5\times1.5~\mathrm{cm}^3$ BGO & 1.1\% \\
        Crystal non-uniformity -- $12\times2\times2~\mathrm{cm}^3$ BGO     & 0.4\% \\
        Crystal non-uniformity -- $12\times2\times2~\mathrm{cm}^3$ BSO     & 1.6\% \\
        \bottomrule
    \end{tabular}
\end{table}



\section{Discussions}
\label{sec:discussions}

Through dedicated beam tests of crystal--SiPM assemblies, the non-linear response of high-pixel-density SiPMs with pixel pitches of 6--10~$\mu$m to scintillation light has been quantified for the first time under realistic beam conditions. Despite these achievements, several limitations of the present study should be noted:
Measure the gain at low temperature, then infer the gain at the target temperature using gain measurements taken at multiple temperatures.
\begin{itemize}
    \item The charge-injection calibration of the pre-amplifiers relied on signals generated by a waveform generator, which may not perfectly reproduce the temporal and spectral characteristics of actual SiPM signals.
    \item For some SiPMs, the signal-to-noise ratio at the beam-test operating temperature were insufficient to allowdirect gain calibration. Consequently, the gains were measured at lower temperatures and then extrapolated to the beam-test temperatures. This procedure introduces additional systematic uncertainty.
    \item In the SiPM readout circuit, the cathode bias line includes a series resistor of 10~k$\Omega$, and the anode signal is AC-coupled. These design features can lead to incomplete bias recovery for closely spaced signals, potentially reducing the amplitude of successive pulses.
    \item The performance of the NDL SiPMs is significantly worse than expected, which may be related to device-specific effects, such as a fraction of non-functional pixels or gain variations induced by changes in pixel capacitance under large signal conditions. Further dedicated studies are required to clarify the underlying mechanisms.
\end{itemize}


\section{Summary and outlook}
\label{sec:summary}

A dedicated beam test was performed to investigate the non-linear response of high-pixel-density SiPMs coupled to scintillating crystals under conditions relevant for homogeneous electromagnetic calorimetry at future Higgs factories. By employing a dual-end readout configuration with an attenuated reference SiPM, together with a tungsten pre-shower and variable incident angles, energy depositions of larger than 100~GeV were achieved in BGO and BSO crystal bars. This approach enabled a precise calibration of the number of photoelectrons over an extremely wide dynamic range far exceeding that accessible in conventional laboratory measurements.

The results show that the scintillation duration of BGO and BSO plays a crucial role in mitigating SiPM saturation through pixel recovery effects, effectively extending the usable dynamic range beyond the nominal pixel count. For BGO crystals, the non-linearity of Hamamatsu SiPMs with 10~$\mu$m pixel pitch remains at the level of about 20\% at $5\times10^{5}$ photoelectrons, while faster scintillators such as BSO exhibit more pronounced saturation effects. Comparisons between different crystal lengths indicate that the intrinsic scintillation decay time dominates the non-linear behavior, whereas photon transport effects associated with crystal geometry have a smaller impact. The performance of the NDL SiPMs appears to be worse than expected, and the underlying reason remains to be investigated.

These measurements constitute one of the first systematic experimental studies of SiPM nonlinearity driven by scintillation light in a beam environment. More importantly, this study aims to provide a research method that can be extended to other types of SiPM-based detectors, specifically for quantitatively evaluating and correcting SiPM response nonlinearity. Looking forward, further improvements can be achieved by refining the front-end electronics design, extending measurements to additional SiPM technologies, and improving the modeling of pixel recovery and saturation effects in simulations. Together, these efforts will contribute to the optimization of SiPM-based calorimeters and support their application in future experiments.


\section*{Acknowledgments}

The authors would like to express their sincere gratitude to fruitful discussions in the CEPC calorimeter working group, the CALICE and DRD6 collaborations. In the beamtest campaign mentioned in this paper, the authors received tremendous support from the CERN SPS testbeam facility and also financial support from the EuroLabs funding support on the Transnational Access.

This work was supported by the following funding agencies: National Key R\&D Program of China (Grant No.: 2023YFA1606904 and 2023YFA1606900), National Natural Science Foundation of China (Grant No.: 12150006), Shanghai Pilot Program for Basic Research—Shanghai Jiao Tong University (Grant No.: 21TQ1400209), and National Center for High-Level Talent Training in Mathematics, Physics, Chemistry, and Biology.


\begin{thebibliography}{00}


\bibitem{CEPCTDR-Acc}
The CEPC Study Group,
CEPC Technical Design Report: Accelerator.
Radiat. Detect. Technol. Methods \textbf{8}, 1-1105 (2024).
doi:10.1007/s41605-024-00463-y.

\bibitem{CEPCTDR-Det}
The CEPC Study Group,
CEPC Technical Design Report -- Reference Detector. 
arXiv:2510.05260 [physics.hep-ex].

\bibitem{Liu_2020}
Y.~Liu, J.~Jiang, Y.~Wang,
High-granularity crystal calorimetry: conceptual designs and first studies.
JINST \textbf{15}, C04056 (2020).
doi:10.1088/1748-0221/15/04/C04056.

\bibitem{instruments6030040}
B.~Qi, Y.~Liu,
R\&D of a Novel High Granularity Crystal Electromagnetic Calorimeter.
Instruments \textbf{6}, 40 (2022).
doi:10.3390/instruments6030040.

\bibitem{qi2026}
B.~Qi, F.~Guo, Y.~Liu, etc.,
Conceptual Design of a Novel Highly Granular Crystal Electromagnetic Calorimeter for Future Higgs Factories.
arXiv:2602.09836 [physics.ins-det]

\bibitem{PIEMONTE20192}
C.~Piemonte and A.~Gola,
Overview on the main parameters and technology of modern Silicon Photomultipliers.
Nucl. Instrum. Methods A \textbf{926}, 2-15 (2019).
doi:10.1016/j.nima.2018.11.119.

\bibitem{KLANNER201936}
R.~Klanner,
Characterisation of SiPMs.
Nucl. Instrum. Methods A \textbf{926}, 35-56 (2019).
doi:10.1016/j.nima.2018.11.083.

\bibitem{Geant4}
Geant4.
https://geant4.web.cern.ch.

\bibitem{PDG}
Paticle Data Group.
https://pdg.lbl.gov/.

\bibitem{JI2014143}
Z.~Ji, H.~Ni, L.~Yuan, et al.,
Investigation of optical transmittance and light response uniformity of 600-mm-long BGO crystals.
Nucl. Instrum. Methods A \textbf{753}, 143-148 (2014).
doi:10.1016/j.nima.2014.03.056.

\bibitem{ISHII2002201}
M.~Ishii, et al.,
Development of BSO (Bi$_4$Si$_3$O$_{12}$) crystal for radiation detector.
Opt. Mater. \textbf{19}, 201-212 (2002).
doi:10.1016/S0925-3467(01)00220-8.

\bibitem{Hamamatsu}
Hamamatsu,
https://www.hamamatsu.com.cn.

\bibitem{NDL}
Novel Device Laboratory,
http://www.ndl-sipm.net.

\bibitem{Tang_2023}
J.~Tang, W.~Wu, L.~Li, et al.
A fast tunable driver of light source for the TRIDENT Pathfinder experiment.
JINST \textbf{18}, T08001 (2023).
doi:10.1088/1748-0221/18/08/T08001.

\bibitem{Pico}
Pico Technology,
https://www.picotech.com/products/oscilloscope.

\bibitem{Zhao:2024ohy}
Z.~Zhao, B.~Qi, Y.~Liu, etc.,
Studies on Dynamic Range of SiPMs with High Pixel Densities.
EPJ Web Conf. \textbf{320}, 00061 (2025).
doi:10.1051/epjconf/202532000061.

\end{thebibliography}



\end{document}